%% file: bare_jrnl.tex
\documentclass[journal,compsoc,screen]{IEEEtran}
\input{meta/lstcustom.tex}
\input{meta/package.tex}

\input{meta/algorithmstyle}

\ifCLASSINFOpdf
\else
\fi
\begin{document}
%
\title{A Systematical Study on Application Performance Management Libraries for Apps}
%
%

\author{Yutian~Tang,
        Haoyu~Wang,
        Xian Zhan, 
        Xiapu~Luo,
        Yajin~Zhou,\\ 
        Hao~Zhou,
        Qiben~Yan,
        Yulei~Sui,
        Jacky~Keung
\thanks{Y. Tang is with ShanghaiTech University, China}
\thanks{H. Wang is with Beijing University of Posts and Telecommunications, China}
\thanks{X. Zhan, H. Zhou, and X. Luo are with the Hong Kong Polytechnic University, Hong Kong SAR, China.}
\thanks{Y. Zhou is with Zhejiang University, China}
\thanks{Q. Yan is with Michigan State University, USA}
\thanks{Y. Sui is with University of Technology Sydney, Australia}
\thanks{J. Keung is with the City University of Hong Kong, Hong Kong SAR, China}
\thanks{Y. Tang (csytang@ieee.org) and X. Luo (csxluo@comp.polyu.edu.hk) are the corresponding authors;}}

%
%

\markboth{Journal of \LaTeX\ Class Files,~Vol.~14, No.~8, August~2015}%
{Shell \MakeLowercase{\textit{et al.}}: Bare Demo of IEEEtran.cls for IEEE Journals}
%



\IEEEtitleabstractindextext{
\input{sections/abstract}

}

\maketitle


%

\IEEEpeerreviewmaketitle

\input{sections/introduction}
\input{sections/background}

\input{sections/apm}
\input{sections/methodolgy}
\input{sections/empiricalstudy}
\input{sections/lessonslearned}
\input{sections/relatedwork}
\input{sections/conclusion}



%
\bibliographystyle{IEEEtran}
\bibliography{ref/reference}

\balance



\end{document}

%% file: meta/lstcustom.tex
\usepackage{color}
\usepackage{etoolbox}
\usepackage{listings}
\usepackage[T1]{fontenc}

\usepackage[scaled=0.8]{beramono}

\newcommand{\lstfontfamily}{\ttfamily}

\definecolor{darkviolet}{rgb}{0.5,0,0.4}
\definecolor{darkgreen}{rgb}{0,0.4,0.2} 
\definecolor{darkblue}{rgb}{0.1,0.1,0.9}
\definecolor{darkgrey}{rgb}{0.5,0.5,0.5}
\definecolor{lightblue}{rgb}{0.4,0.4,1}

\definecolor{stringColor}{rgb}{0.16,0.00,1.00}
\definecolor{annotationColor}{rgb}{0.39,0.39,0.39}
\definecolor{keywordColor}{rgb}{0.50,0.00,0.33}
\definecolor{commentColor}{rgb}{0.25,0.50,0.37}
\definecolor{javadocColor}{rgb}{0.25,0.37,0.75}
\definecolor{jTagColor}{rgb}{0.50,0.62,0.75}
\definecolor{eTagColor}{rgb}{0.50,0.62,0.75}
\definecolor{lineNumberColor}{rgb}{0.47,0.47,0.47}

\def\jTags{@author, @deprecated, @exception, @param, @return, @see, @serial, @serialData, @serialField, @since, @throws, @version}

\def\jAnnotations{
    classoffset=1,
    morekeywords={@Override, @Deperecated, @SuppressWarnings, @Retention, @Documented, @Target, @Inherited},
    keywordstyle=\color{annotationColor},
    classoffset=0
}

\def\eTags{FIXME, TODO, XXX}

\newrobustcmd{\markupJavadocs}[1]{%
\edef\mytok{\the\lst@token}%
{\color{javadocColor}%
\expandafter\docsvlist\expandafter{\jTags}%
\expandafter\docsvlist\expandafter{\eTags}%
#1}%
}%

\newrobustcmd{\markupComments}[1]{%
\edef\mytok{\the\lst@token}%
{\color{commentColor}%
\expandafter\docsvlist\expandafter{\eTags}#1}%
}%

\lstdefinestyle{eclipse}{
  basicstyle={\lstfontfamily},
  emphstyle=\bfseries,
  keywordstyle=\color{keywordColor}\bfseries,
  commentstyle=\markupComments,
  stringstyle=\color{stringColor},
  numberstyle=\color{lineNumberColor}\lstfontfamily,
  morecomment=[s][\markupJavadocs]{/**}{*/}, 
  showstringspaces=false,
  numbers=left,
}

\lstdefinestyle{black}{
  basicstyle=\small\lstfontfamily,
  numbers=left,
  columns=fullflexible,
  breaklines=true,
  mathescape=true,
  escapechar=\#,
  tabsize=4,
  frame=lines,
  showstringspaces=false
}

\lstdefinestyle{seminar}{
  basicstyle=\small\ttfamily,
  numbers=left,
  breaklines=true,
  mathescape=true,
  escapechar=\#,
  tabsize=4,
  showstringspaces=false
}

\expandafter\lstset\expandafter{\jAnnotations}

%% file: meta/package.tex
\usepackage{hyperref}
\usepackage{mdframed}
\usepackage{lipsum}
\usepackage{amsmath}
\usepackage{amssymb}
\usepackage{amsfonts}
\usepackage[font=small]{caption}
\usepackage{ifpdf}
\usepackage[final]{graphicx}
\usepackage{todonotes}
\usepackage{booktabs} 
\usepackage{url}
\usepackage{amssymb}
\usepackage{pifont}
\usepackage{color}
\usepackage{tcolorbox}
\usepackage{balance}
\usepackage{makecell}
\usepackage{verbatim}
\usepackage{hyperref}

\usepackage{algorithmic}
\usepackage{array}

\usepackage{caption}
\usepackage{multirow}
\usepackage{float}
\usepackage{mwe}
\newcommand{\code}[1]{\texttt{#1}}

\usepackage{url}
\usepackage{cite}
\usepackage{todonotes}
\newcommand{\cmark}{\ding{51}}%
\newcommand{\xmark}{\ding{55}}%

\usepackage{subfigure}
\usepackage{tikz}
\usepackage{xcolor}

\usepackage{booktabs,caption,fixltx2e}
\usepackage[flushleft]{threeparttable}

\hypersetup{
 colorlinks=true,
 linkcolor=blue,
 citecolor=blue,
 filecolor=black,
 urlcolor=black,
 pdfauthor = {},
 pdftitle = {},
 pdfkeywords = {}
}

%% file: meta/algorithmstyle.tex
\usepackage{algorithm2e}
\usepackage{xcolor}

\newcommand{\xvar}[1]{\textsf{#1}}

\newcommand{\xvbox}[2]{\makebox[#1][l]{#2}}

\SetAlFnt{\footnotesize}

\SetAlCapSty{xAlCapSty}

\SetCommentSty{xCommentSty}

\SetNlSty{mynlfont}{}{} 

\LinesNumbered

\SetSideCommentRight

\DontPrintSemicolon

\RestyleAlgo{algoruled}

%% file: sections/abstract.tex
\begin{abstract}


Being able to automatically detect the performance issues in apps can significantly improve apps' quality as well as having a positive influence on user satisfaction. \underline{A}pplication \underline{P}erformance \underline{M}anagement (APM) libraries are used to locate the apps' performance bottleneck, monitor their behaviors at runtime, and identify potential security risks. 
Although app developers have been exploiting application performance management (APM) tools to capture these potential performance issues, most of them do not fully understand the internals of these APM tools and the effect on their apps. 
To fill this gap, in this paper, we conduct the first systematic study on APMs for apps by scrutinizing 25 widely-used APMs for Android apps and develop a framework named APMHunter for exploring the usage of APMs in Android apps.
Using APMHunter, we conduct a large-scale empirical study on 500,000 Android apps to explore the usage patterns of APMs and discover the potential misuses of APMs. We obtain two major findings: 1) some APMs still employ deprecated permissions and approaches, which makes APMs fail to perform as expected; 2) inappropriate use of APMs can cause privacy leaks. Thus, our study suggests that both APM vendors and developers should design and use APMs scrupulously.
\end{abstract}

%% file: sections/introduction.tex
\section{Introduction}\label{sec:introduction}
Android is dominating the market for smartphone operating systems today. There are around 3 million apps in the Google Play store according to a report from AppBrain~\cite{AppBrain}. With high smartphone penetration, mobile apps have become indispensable to billions of users. The runtime performance of an app can significantly affect its user experience. Thus, more and more Android developers tend to employ application performance management (APM) tools to cope with performance bottlenecks~\cite{Yao:2018, Karami:2013}. 


APMS can be used in desktop and mobile applications, such as cloud applications, web applications, and mobile apps. APMs assist developers in locating the potential performance limitations in applications~\cite{Ahmed:2016,Heger:2017}. However, developers may not have profound understandings of the APMs' functionalities~\cite{Heger:2017, Yao:2018, Trubiani:2018} as most of them are commercial (closed source) products. 

Besides, inappropriate use of APMs can cause security issues for apps. For example, the regular logging frameworks (e.g., android.uti.log) allow developers to write custom logs. In most APMs, they also provide such functionality to developers. Since the information recorded in logs is determined by developers, developers may collect sensitive data at runtime. Demystifying the design of APMs and exploring the usage practices of APMs can benefit all stakeholders, including APM vendors, developers, and app users. To assist developers in understanding the functionalities of APMs, in this paper, we conduct a systematical study on APMs for Android apps. Besides providing insightful understandings of APMs, we  reveal seven design defects in existing APMs (see Sec. \ref{sec:apm}) which can be used to improve existing APMs.

\noindent\textbf{Motivation.} Existing studies on APMs mainly explain how to use the data collected by APMs to diagnose or locate the problems (e.g., bugs) in a program \cite{ Ahmed:2016, Heger:2017,Willnecker:2015}. For instance, Ahmed et al. \cite{Ahmed:2016} discussed whether APMs can detect the performance regressions (e.g., excessive memory usage, high CPU utilization, and inefficient database queries) for web applications. They conducted an empirical study on three commercial APMs and an open-source APM to examine whether these APMs can help developers diagnose the performance regressions. Trubiani et al. \cite{Trubiani:2018} leveraged the Kieker APM to detect performance anti-patterns in load testing. Streitz et al. \cite{Streitz:2018} presented how SAP company employs APMs for performance prediction. Heger et al. reported~\cite{Heger:2017} the activities and key concepts in an APM from the data perspective, such as  data collection, data processing, data interpretation. 

Existing research does not reveal the implementation details of APMs. As a result, developers use APMs as black-box tools. They only have a general but vague idea about these APMs instead of an insightful understanding. Apart from discussing the implementations of APMs, we also explore potential security issues, including whether there are defects in APMs, whether APMs actively collect data from users, whether APMs can be exploited by attackers. To fill this gap, we conduct a thorough study on 25 Android-oriented APMs and demystify their functionalities. Our research can enlighten developers about APMs with the points they usually ignore.



\noindent\textbf{Major Extension.} This paper is an extension of \cite{Tang:2019}, which was published in the 34th IEEE/ACM International Conference on Automated Software Engineering (ASE 2019). The major extensions include:

\noindent $\bullet$ We provide additional background knowledge on two instrumentation approaches, which include manual instrumentation and auto-instrumentation. We compare them and present the differences and their suitability;

\noindent $\bullet$ We conduct additional analysis on commercial APMs (see Sec. \ref{sec:apm}). To be exact, we present the limitations and drawbacks of existing APMs (see Sec. \ref{subsec:limitations-drawbacks}). Meanwhile, we discuss whether existing APMs are qualified for monitoring apps' runtime performance (see Sec. \ref{subsec:confront-anti-patterns});

\noindent $\bullet$ We define and propose a new framework named \code{APMHunter}, which can automatically detect the APM usages in the apps (see. Sec. \ref{sec:methodolgy}). \code{APMHunter} contains three modules: \textit{APM API identification} (i.e., for detecting the usage of APMs), \textit{static analysis} (i.e., app analysis framework), and \textit{usage identification} (i.e.,  locating and reasoning the usage of APMs in apps). It is worth mentioning that our method is obfuscation resilient so that it can handle both obfuscated and none-obfuscated apps. Furthermore, we also design a set of experiments to evaluate its accuracy (see Sec. \ref{subsec:evaluation-apmhunter});

\noindent $\bullet$ We conduct an additional empirical study on apps (see Sec. \ref{sec:empiricalstudy}). Compared with the conference paper, we extend the two research questions (RQ1, RQ4), including (1) we rank APMs based on their popularity. We find that developers overwhelmingly choose commercial APMs rather than open-source APMs. We present 4 possible concerns for developers to use commercial APMs (see RQ1); and (2) comparing with the conference version, we extend our tool to support detecting privacy leaks from user inputs. We conduct additional experiments on detecting privacy leaks from user inputs to APMs (see RQ4). We also answer three new research questions (RQ2, RQ3, RQ5), including (1) as most APMs provide built-in logging functions for developers, we analyze the intention and consequences of using logging frameworks and APMs respectively (see RQ2); (2) we investigate the consequences of using multiple APMs in one app (see RQ3); and (3) we evaluate the performance overhead introduced by using an APM in an app. The corresponding results can assist developers in deciding whether to use APMs in their apps (see RQ5); and

\noindent $\bullet$ We add a new section (see Sec. \ref{sec:lessons-learned}) to provide guidance for stakeholders. For APM vendors, we offer four suggestions for them to develop APMs, which cover the vulnerabilities and limitations we found in the existing commercial APMs. For app developers, we present two suggestions for using APMs in their apps. 

\noindent\textbf{Contribution.} In summary, our key contributions includes:

\noindent$\bullet $ To the best of our knowledge, this is the \emph{first} work that conducts a comprehensive study to demystify the functionalities of Android-oriented APMs. We select 9 major functions in APMs and introduce the implementation details of these APMs, i.e., how these functions are implemented. We reveal 7 design defects in these APMs (see Sec. \ref{sec:apm}); 

\noindent$\bullet $ We develop a prototype named APMHunter to automatically identify APMs used in an app, record APM usages, and report privacy leaks from the app to APMs. Moreover, APMHunter can process both obfuscated apps and non-obfuscated apps (see Sec. \ref{sec:methodolgy}); and

\noindent$\bullet $ We conduct a large-scale empirical study on 500,000 Android apps fetched from Google Play to explore how APMs are used in apps, what the side-effects are introduced by APMs, and whether user privacy is leaked to APMs. We find that 23,397 apps will collect sensitive data from users through APMs (see Sec.\ref{sec:empiricalstudy});  


\noindent\textbf{Skeleton.} The rest of this paper is organized as follows: we present the background and usage of APMs in Sec.\ref{sec:background}, and elaborate the functionalities of these APMs in Sec. \ref{sec:apm}. To explore the usage of APMs in an app, we develop a tool named \code{APMHunter}, whose design and implementation are described in Sec. \ref{sec:methodolgy}. Sec. \ref{sec:empiricalstudy} presents the details of our empirical study. Furthermore, we discuss the lessons learned and several keys issues of APMs in Sec. \ref{sec:lessons-learned}. After presenting the related work in Sec. \ref{sec:relatedwork}, we conclude the paper in Sec. \ref{sec:conclusion}.

\noindent\textbf{Data Availability.} The experimental data and our tool APMHunter are available at:
\url{https://sites.google.com/view/systematical-apm-study}.

%% file: sections/background.tex
\section{Preliminary}\label{sec:background}

This section introduces the motivation for using APMs in apps and how to embed an APM into an app.

\subsection{The Need of APMs} 
Developers debug their apps locally before APMs become popular. They cannot know how apps work in different environments, as it can be hard for developers to test an app on all types of devices from all devices vendors (e.g., Blackberry, Google Pixel, Huawei, SAMSUNG). 
Moreover, through local debugging, developers cannot collect the runtime performance of their apps once the apps are released. To enable tracking runtime performance of apps, APMs play the key role. By integrating APMs into apps, developers can collect the runtime performance of apps even the apps have been released and distributed.

\subsection{How to Use APMs in Apps?}\label{subsec:apm-integration-procedure}
To use an APM, a developer usually takes the following three steps:

\noindent$\bullet$ First, the developer registers an account for the APM. The account is used to login to the APM console to view the performance of the monitored app. Then, the developer needs to register an app with the APM vendor to obtain a unique ID for tracking the app;

\noindent$\bullet$ Second, the developer downloads the APM Standard Development Kit (SDK) and integrates it into the apps. 

\noindent$\bullet$ Last, the developer publishes his app via app stores (e.g., Google Play, Amazon) and collects runtime performance of the app through the APM. The data is then displayed on the APM's console.

\subsection{Manual Integration vs. Automatic Integration}
\label{sec:manual_auto_integration}
\begin{figure}[htpb]
    \centering
    \includegraphics[width=0.5\textwidth]{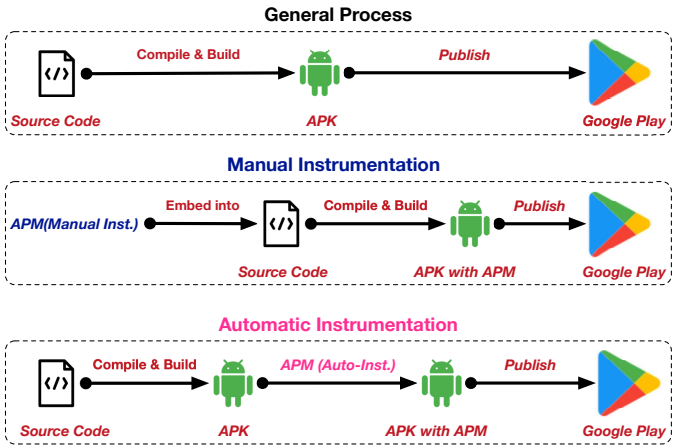}
    \caption{Workflow for Integrating APMs}
    \label{fig:apm-embedding-timeline}
\end{figure}

Existing APMs support two ways of integration: manual integration and automatic instrumentation. The differences between the manual integration and automatic integration are listed as follows (also see Fig. \ref{fig:apm-embedding-timeline}).

\noindent$\bullet$ \textit{Manual integration} requires developers to manually integrate APM SDKs into apps. As illustrated in Fig. \ref{fig:apm-embedding-timeline}, developers add an APM SDK as a dependency in the source code. Once the source code is built, the APM is embedded in the APK (bytecode of the app). The procedure is the same as using a third-party library in an app. Most APMs (e.g., UMeng~\cite{UMeng}, Flurry~\cite{Flurry}) adopt this strategy. 

\noindent$\bullet$ \textit{Automatic integration} works on app binaries rather than source code. APM vendors provide scripts for automatically injecting APMs (e.g., CA Mobile~\cite{CA}) into app binaries (APKs).
The manual integration is suitable for apps whose source code is available, whereas the automatic integration usually aims at binaries.

There is no evidence on which approach is better. But it is worth mentioning that attackers can repackage victim apps \cite{ShaoACSAC14} by leveraging APMs which support automatic integration. For example, it can be hard for attackers to obtain the source code of the Facebook app but the app is available on Google Play. Thus, the attackers can instrument the Facebook app through APMs that support automatic integration, and then publish the repackaged app on some third-party app markets to collect users' data. We compare the usage of these two strategies and find that developers overwhelmingly choose APMs with manual integration. The corresponding data and detailed explanation are presented in Sec. \ref{subsec:popularity}.

%% file: sections/apm.tex
\section{Application Performance Management}\label{sec:apm}

\subsection{Selection Criteria}

Since Android has a prominent position in the entire mobile eco-system, in this paper, we focus on APMs targeting on Android apps. Since previous studies on APMs do not offer a list of Android-oriented APMs\cite{Trubiani:2018,Ahmed:2016,Willnecker:2015,Yao:2018}, we select the target APMs through the following steps: (1) As APMs are considered as third-party libraries, we first crawl the candidate third-party libraries belonging to the category of development tools from AppBrain. Specifically, we crawl the metadata of these libraries, which include descriptions, the market share, tags (e.g., crash reporting, open source), and the official site~\cite{AppBrainThirdParty}; (2) To select the target APMs from these candidates, we carefully define the following criteria by referencing \cite{appdevsurvey1}:

\noindent$\bullet$ The APM must be able to monitor Android apps;

\noindent$\bullet$ The APM can support at least 3 common functionalities. The common functionalities include capturing crashes, correlating server performance, logging, tracking user behaviors, and capturing ANR. 

As a result, we obtain 25 APMs that meet the criteria as shown in Table \ref{tab:apm-programmingpattern}, including 20 commercial APMs and 5 open-source APMs. The statistics from AppBrain \cite{AppBrainThirdParty} show that the selected APMs hold over a 90\% market share \cite{AppBrainThirdParty} of the apps that use APMs. Note that, unlike other APMs, Google Firebase is a collection of libraries, including ad analysis, A/B testing, remote configuration, cloud messaging, and performance diagnosis. In this paper, we only focus on Firebase Analytics, because it allows developers to track users' behavior and capture crashes at runtime. Table~\ref{tab:apm-programmingpattern} lists the functions supported by these APMs and their integration methods. In the coming sections, we introduce how these functions are implemented by reverse-engineering all these APMs.



\begin{table*}[htpb]
\centering
\caption{APM Libraries Studied}
\scalebox{0.91}{
\begin{threeparttable}
\begin{tabular}{c|c|c|c|c|c|c|c|c|c|c}
\hline
\multirow{2}{*}{\textbf{Lib}} & \multicolumn{8}{c|}{\textbf{Monitoring}}                                                                      & \multirow{2}{*}{\textbf{Free$-$Pay}} & \multirow{2}{*}{\textbf{Integration}} \\ \cline{2-9}
                              & \textbf{Crash(Java-Native)} & \textbf{Network} & \textbf{ANR} & \textbf{CPU} & \textbf{Mem.} & \textbf{ToP} & \textbf{Log} &\textbf{Event Track.} &                     &                                         \\ \hline
  Tingyun \cite{TingYun}  & \cmark $-$ \xmark & \cmark & \cmark  & \cmark &     \cmark & \cmark  & \cmark & \cmark & \cmark $-$\cmark &  source  code     \\ \hline
 BaiduAPM \cite{MTJBaidu}  & \cmark $-$\cmark  &  \cmark    &     \cmark   &  \cmark  &   \cmark    &   \xmark   &  \cmark  &     \cmark &    \cmark $-$\cmark  &  source code   \\ \hline
  UMeng \cite{UMeng}  &  \cmark $-$\cmark  &    \xmark              &  \xmark     &    \xmark    &  \xmark   &  \xmark  & \cmark    &   \cmark &\cmark $-$\cmark  &   source code   \\ \hline
  Mobile Tencent \cite{MobileTencentAnalytics}  &   \cmark $-$\cmark     &      \cmark   &    \xmark       &    \cmark    &     \cmark &  \xmark  &  \cmark   & \xmark & \cmark$-$\xmark     &  source code \\ \hline
 OpenInstall \cite{OpenInstall} &  \xmark $-$\xmark &             \cmark  &  \xmark &  \xmark  &   \xmark  &  \xmark   &  \cmark & \cmark & \xmark $-$\cmark &   source code     \\ \hline
 New Relic \cite{NewRelic} &  \cmark $-$\xmark  &  \cmark   &     \xmark    &     \cmark    &   \cmark    &   \xmark    &    \cmark   & \cmark   &   \xmark $-$\cmark & source code  \\ \hline
 App Dynamics \cite{AppDynamics} &    \cmark$-$ \xmark & \cmark &   \xmark   &   \xmark    &     \xmark     & \xmark    & \cmark &  \cmark &    \xmark $-$\cmark  &   source code     \\ \hline
 OneAPM \cite{OneAPM} &  \cmark$-$ \cmark  & \cmark      & \cmark &   \cmark   &  \cmark     &  \xmark    &  \xmark       & \xmark    &   \xmark $-$\cmark &       source code    \\ \hline
 GrowingIO \cite{GrowingIO} &  \xmark $-$\xmark  &  \cmark   &  \xmark    &  \xmark   &  \xmark   &  \cmark   &  \xmark &  \cmark    & \xmark $-$\cmark  &      source code  \\ \hline
 Google Firebase \cite{GoogleFirebase} &  \cmark$-$\cmark       &   \cmark        & \xmark  &   \xmark  &  \xmark   &  \xmark &  \cmark  &   \cmark    &  \cmark $-$\cmark &   source code  \\ \hline
 Dynatrace \cite{Dynatrace} &   \cmark$-$  \xmark  &  \cmark        &     \xmark       & \cmark       &    \cmark      & \xmark        &  \cmark    & \cmark   &  \xmark $-$\cmark  &   source code      \\ \hline
 Site24$\times$7 \cite{Site24-7} &  \xmark $-$  \xmark   &  \cmark   &   \xmark        &  \cmark   &   \cmark   &  \xmark       &   \xmark    &   \cmark  &    \xmark $-$\cmark    &  source code        \\ \hline
 %
 AppPulse Mobile\cite{AppPulse}  &  \cmark$-$  \xmark    & \cmark   &  \xmark     &   \xmark   & \cmark       &  \xmark  &  \xmark  &  \cmark   & \xmark$-$\cmark  &    bytecode         \\ \hline
 CA Mobile\cite{CA}  & \cmark$-$  \xmark   & \cmark   & \cmark       &    \cmark     & \cmark     &   \xmark  &  \xmark     &   \cmark   &  \xmark$-$\cmark    &      bytecode                                   \\ \hline
\makecell{Apteligent}\cite{Apteligent}  &  \cmark $-$ \cmark   & \cmark  &    \xmark    &    \cmark     &  \cmark        &  \xmark &  \cmark    &    \cmark &    \xmark $-$\cmark  &   source code   \\ \hline
 %
 %
 %
 
 Flurry \cite{Flurry} & \cmark $-$\cmark &   \xmark &   \xmark &  \xmark   &   \xmark &   \xmark   & \cmark &   \cmark  &   \xmark $-$\cmark  &   source code       \\ \hline
 AppsFlyer \cite{AppsFlyer}  &   \xmark  $-$ \xmark    &  \xmark &     \xmark &  \xmark &   \xmark   &  \xmark &  \cmark    &   \cmark &   \xmark $-$\cmark   & source code \\ \hline
  %
   %
 Yandex Metrica \cite{AppMetrica}  &  \cmark $-$ \cmark   &   \cmark  & \xmark  &  \xmark  &  \xmark    &   \xmark   & \cmark      &  \cmark  &  \xmark $-$ \cmark &  source code   \\ \hline
 Adjust \cite{Adjust} & \cmark $-$ \xmark  &   \xmark  &  \xmark &      \xmark    &    \xmark   &   \xmark    &      \cmark & \cmark & \xmark $-$ \cmark & source code   \\ \hline
 Ironsource\cite{Ironsource}  &  \cmark$-$\xmark   &  \cmark   & \xmark      &    \xmark   &   \xmark & \xmark    &  \cmark   &    \cmark     &   \xmark $-$ \cmark  &   source code   \\ \hline
 Countly \cite{Countly-android} &   \cmark $-$ \xmark  &     \cmark   & \xmark  &  \xmark  &   \xmark    &    \xmark &   \cmark  &   \cmark    &  open source    &    source code  \\ \hline
 Sentry \cite{Sentry} &   \cmark $-$ \xmark  &   \xmark   &    \cmark    &\cmark   &  \cmark & \cmark & \cmark &  \cmark   &  open source    &  source code \\ \hline
AndroidGodEye \cite{AndroidGodEye} &     \cmark $-$ \xmark           &   \cmark    & \cmark  & \cmark &  \cmark  &  \xmark     &  \cmark  &    \xmark   &  open source    & source code  \\ \hline
BlackCancary \cite{BlackCancary} & \xmark $-$ \xmark  &  \xmark  & \cmark     &   \cmark &  \cmark  &  \cmark  &    \xmark     &  \xmark   &  open source    &    source code \\ \hline
ArgusAPM \cite{ArgusAPM} & \cmark $-$ \xmark  &  \cmark   & \cmark     &   \cmark &  \cmark  &  \cmark  &  \cmark      &  \cmark   &  open source    &    source code \\ \hline

\end{tabular}
\begin{tablenotes}
    \item Network: network diagnosis; ANR: Android Not Responding; CPU: CPU utilization; Mem.: Memory usage; ToP: Time on page; Log: customized logging.
\end{tablenotes}
\end{threeparttable}
}
\label{tab:apm-programmingpattern}
\end{table*}

\subsection{Capturing Crash in Java Code}
When a crash happens in Java code, an APM captures the uncaught \code{exception} and records the execution trace for the crash, which provides more information about how the exception is triggered. If an exception is not captured by any \code{try-catch-finally} block, it is then treated as an uncaught exception, which causes the app to crash. The APMs that can capture such crashes follow the same procedure: they register an uncaught exception handler (using \code{Thread.setDefaultUncaughtExceptionHandler}), a customized instance of \code{Thread.UncaughtExceptionHandler}, to the current thread. When an uncaught exception occurs, the uncaught exception handler implemented by APMs captures the exception.

\begin{tcolorbox}[boxrule=1pt,boxsep=1pt,left=2pt,right=2pt,top=2pt,bottom=2pt]
The main limitation is that using setDefaultUncaughtExceptionHandler can update the \code{UncaughtExceptionHandler} for the Android framework. If an app uses two APMs, only the last initialized APM can capture uncaught exceptions, because only one \code{UncaughtExceptionHandler} can be defined as the default handler (more details are introduced in \textsection \ref{sec:empiricalstudy}).
\end{tcolorbox} 

\subsection{Capturing Crash in Native Code}
Crashes can also happen in native code (C/C++) of an Android app. APMs handle native crashes with the following steps: installing a signal handler, extracting the stack traces, and building the symbol files.


\noindent $\bullet$ \underline{Installing a signal handler.} When crashes occur in native code, an error signal is generated \cite{Ritchie:1975c,GNU}. APMs can capture this error signal by using the method \code{sigaction} to register a signal handler.

\noindent $\bullet$ \underline{Extracting the stack traces.} After receiving the signals, the APM copies the crashed process to a daemon process, which shares an address space with the crashed process. This allows the APM to trace the crashed process. 


\noindent $\bullet$ \underline{Building the symbol files.} APMs locate the start point of the crash to recover the crash point from the program. After that, APMs generate human-readable stack traces and symbol files. The symbol files hold a variety of data, such as function, module, call frame information, which helps with the debugging process.



Interested readers are referred to Google's Breakpad \cite{GoogleBreadPad} framework for details. Breakpad is a mature and open-source library for debugging and analyzing crashes for C/C++ program. It is used in most commercial APMs (e.g., UMeng, Tingyun, Sentry).
%

\subsection{Network Diagnosis}
APMs can also be leveraged for diagnosing the network bottleneck and monitoring network performance. In general, there are two widely adopted solutions: socket based solution and aspect-oriented programming (AOP) \cite{Safonov:2008} based solution. 

\noindent\textbf{Socket Connection Monitoring.} APMs can track the network requests by monitoring the socket in use. Specifically, it can be realized by implementing the \code{SocketImplFactory} interface and then setting the customized \code{SocketImplFactory} as the default. The information about the IP address and port of the target server can be obtained through Java Reflection \cite{Java}.

\begin{figure}[!htpb]
    \centering
    \includegraphics[width=0.5\textwidth]{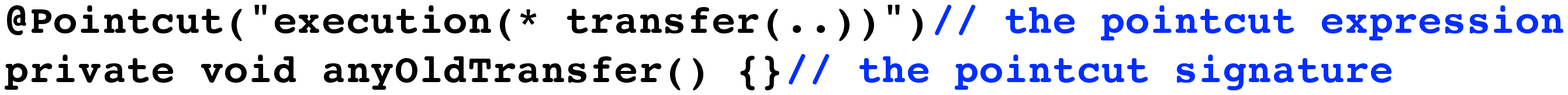}
    \caption{The Structure of A Pointcut in AOP}
    \label{fig:aop-pointcut-structure}
    \vspace{-2ex}
\end{figure}

\begin{figure}[!htpb]
    \centering
    \includegraphics[width=0.5\textwidth]{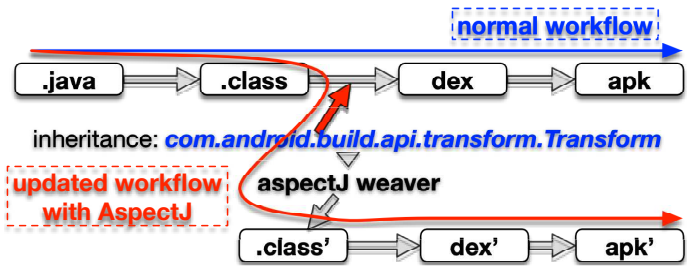}
    \caption{Workflow of AspectJ in APM}
    \label{fig:aop-runtime-weaver}
    \vspace{-2ex}
\end{figure}

\begin{figure}[!htpb]
    \centering
    \includegraphics[width=0.5\textwidth]{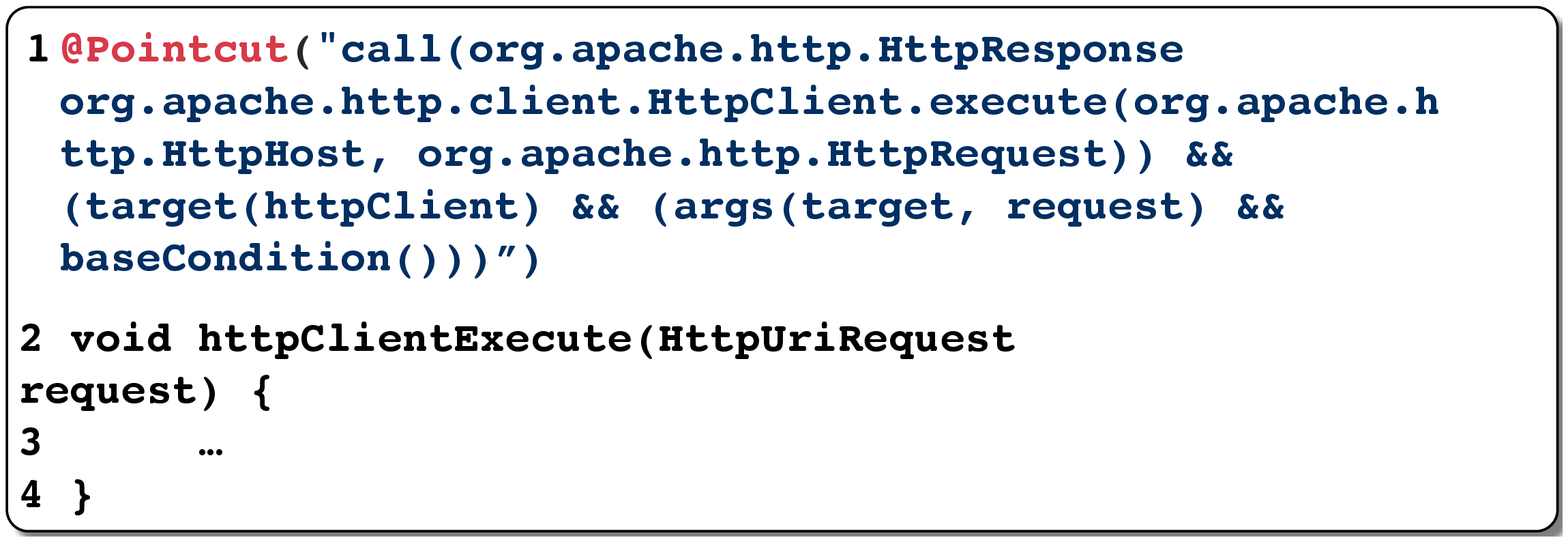}
    \caption{Http Requests Interception based on AOP}
    \label{fig:aop-interceptor}
    \vspace{-1ex}
\end{figure}

\noindent\textbf{Using AOP for Interception}
Another strategy for monitoring and measuring URI requests sent by apps is using AOP. In APMs, the AspectJ is used for implementing AOP \cite{AspectJ}. The workflow of AspectJ in an APM is shown in Fig. \ref{fig:aop-runtime-weaver}. When a developer builds an app with \code{gradle} \cite{Gradle}, the APM can hook the transformation process from \code{classes} to a \code{dex} file by inheriting the \code{Transform} class \cite{Android}. Then, the AspectJ weaver \cite{AspectJ} injects the customized code into the original \code{classes} and then builds the customized \code{dex} file. 

AspectJ allows developers to leverage \code{Pointcut} to implement code injection for runtime monitoring. As shown in Fig. \ref{fig:aop-pointcut-structure}, a pointcut contains two elements: a signature that comprises a name and parameters, and a pointcut expression that determines the method executions to track. Fig. \ref{fig:aop-interceptor} shows an example, where the pointcut designator \code{call} is used to match all method executions whose method signatures are defined in the pointcut expression. The pointcut designator \code{target} can limit matching to join points (execution of methods). Consequently, with all these pointcut designators, APMs can capture network requests at runtime. 


\begin{tcolorbox}[boxrule=1pt,boxsep=1pt,left=2pt,right=2pt,top=2pt,bottom=1pt]
The AOP-based approach relies on \code{Transform}, which only supports the transformation from \code{class} to \code{dex} file with Gradle build. It means that the AOP-based approach is only suitable for apps built with Gradle \cite{Gradle}.
\end{tcolorbox} 
\vspace{-1ex}

\subsection{Analyzing Android Not Response (ANR) Error}
Application Not Responding (ANR) error is another typical performance issue in apps. When a user interface (UI) thread of an app is blocked for a long time, the ANR error is triggered. APMs use the following approaches to track ANR errors:

\noindent\textbf{Solution 1.} APMs can implement a watchdog to detect ANR errors. The watchdog itself is a \code{thread}. It checks the status of main \code{thread} in a periodic way. If the main \code{thread} is frozen for over \code{n} seconds (\code{n} is a predefined by APM vendors), the watchdog reports the ANR error.

\noindent\textbf{Solution 2.} As Android is a message-driven system, system events are scheduled and appended to the message queue. The main thread (a.k.a \code{Looper} thread) is responsible for looping the message queue and handling messages in the message queue continuously. When the \code{Looper} is blocked (ANR error), Android outputs the ANR error into a certain trace file (\code{data/anr/traces.txt}). By rewriting the \code{Looper.getMainLooper().setMessageLogging(Printer printer)} API, APMs can capture the ANR. This is because once the ANR occurs, Android records the ANR error using the default \code{Printer} and writes to the trace file (\code{data/anr/traces.txt}). As APMs override the \code{Printer}, APMs can capture the ANR error.

\begin{tcolorbox}[boxrule=1pt,boxsep=1pt,left=2pt,right=2pt,top=2pt,bottom=2pt]
Compared with Solution 2, Solution 1 has two limitations: (1) the watchdog thread has to keep checking the main thread to capture the ANR error; and (2) it is difficult to set a proper timeout value for the watchdog thread. A small timeout value can cause performance overhead as it frequently checks the main thread. However, a large timeout can make the watchdog fail to report ANR errors promptly.
\end{tcolorbox} 

\subsection{Time-on-page Analysis}
The time-on-page (ToP) analysis aims at monitoring the time spent on a page (e.g., an \code{Activity}). To compute the time-on-page, APM must be able to capture UI display transitions. When a UI display changes, a new page is loaded. It suggests a display transition.

APMs apply \code{Choreographer.FrameCallback.doFrame()} to monitor UI display transitions. In Android, the Choreographer component receives timing pulses from the display, and then it schedules the rendering work for the next display frame \cite{Android}. The callback method \code{doFrame} is automatically invoked by Android when Android starts rendering the next display frame.

\begin{tcolorbox}[boxrule=1pt,boxsep=1pt,left=2pt,right=2pt,top=2pt,bottom=2pt]
The limitation of the ToP analysis is that the Choreographer API is introduced since Android 4.1 (API 16). Thus, it cannot be used for devices with an API level lower than 16.
\end{tcolorbox} 

\subsection{Logging and Tracking}
Developers can employ the logging functions provided by APMs to collect users' execution traces during the runtime. The information recorded with the built-in logging function in APM is sent back to the server. Developers can also leverage APMs to track any concerned event. For example, in New Relic APM, developers can use \code{recordCustomEvent(type,name,attributes)} to record an event at runtime. In practice, developers can use such APIs to collect user behaviors, such as preferences and execution paths.


\subsection{Other supporting functions}\label{subsec:support-functions} 

\noindent\textbf{Memory usage.} Memory usage data can be used to diagnose potential memory leaks in the app. In general, there are three approaches for collecting memory usage data: (1) using the Android API \code{ActivityManager.MemoryInfo}; (2) accessing the system file \code{/proc/meminfo}; and (3) utilizing the Android API \code{ActivityManager.getProcessMemoryInfo}.

The method (1) and (3) can provide memory usage information of the target app. The method (2) provides the memory usage of all processes running in Android. Then, the memory usage information for the target app is then filtered by APMs.

\noindent\textbf{CPU Utilization.} To capture the CPU utilization, APMs obtain the CPU usage by inspecting system files. These system files include \code{/proc/cpuinfo}, \code{/proc/<pid>/stat}, \code{/proc/stat}, and \code{/sys/devices/system/cpu/cpu0}. Since Android 8.0 (API 26), the file \code{/proc/stat} cannot be visited without the root permission. 

\noindent\textbf{Time Consuming.} To compute the time consumed by a code fragment, APMs mainly take advantage of two APIs: \code{currentTimeMillis} and \code{TimeUnit.MILLSECOND}. Both are defined in Java SDK.

\begin{tcolorbox}[boxrule=1pt,boxsep=1pt,left=2pt,right=2pt,top=2pt,bottom=2pt]
There is a compatibility defect in the existing APMs that the file \code{/proc/stat} cannot be visited since Android 8.0 (API 26). APMs cannot collect CPU usage of all active processes with this approach. However, developers can still collect the CPU usage from other system files (e.g, /proc/cpuinfo), which leads to the leak of private data (e.g., CPU family, CPU model) through APM (also reference \cite{Zhou:2013}).
\end{tcolorbox} 

\subsection{Limitations and Drawbacks of APMs}\label{subsec:limitations-drawbacks}

\noindent\textbf{Requesting unnecessary or dangerous permissions.}
Some APMs request permissions that are proven to be deprecated or unnecessary. These permissions include READ\_LOGS, READ\_PHONE\_STATE, GET\_TASK, BLUE\_TOOTH, SYSETM\_ALERT\_WINDOW, and SYSETM\_OVERLAY\_WINDOW. Specifically, the permission READ\_LOGS and GET\_TASK are deprecated. Some permissions should not be granted, such as device state and Bluetooth state. This information usually should not be leaked to developers.
A number of research studies point out that some attacks are strongly related to these permissions \cite{fratantonio17,Au:2012,Michael:2016,Felt:2011}.

\begin{table*}[!htpb]
\centering
\caption{APMs' Capability on Handling Performance Anti-patterns}
\begin{threeparttable}
\begin{tabular}{|c|c|c|c|}
\hline
Anti-pattern & APMs Support & Anti-pattern & APMs Support \\ \hline
(1) GUI lagging   & 1,2,8,14,22,23,24,25   &  (2) Energy leak  &  Nil    \\ \hline
(3) Memory bloat   &  1,2,4,6,8,11,12,13,14,15,22,23,24,25    &  (4) Cyclic/Frequent invo.    &  1,2,4,6,8,11,12,14,15,22,23,24,25   \\ \hline
(5) Expensive callee   &  1,2,4,6,8,11,12,14,15,22,23,24,25    & (6) Loading time  &  1,2,4-15,18,20,21,23,25    \\ \hline
(7) Query local DB  &  Nil    &  (8) UI overdraw \cite{Hecht:2015}  & Nil \\ \hline

\end{tabular}
\begin{tablenotes}
    \item 1:Tingyun; 2:BaiduAPM; 3:UMeng; 4: Mobile Tencent; 5: OpenInstall; 6: New Relic; 7: App Dynamics; 8: OneAPM; 9 GrowingIO; 10: Google Firebase; 11: Dynatrace; 12: Site24 $\times$ 7; 13: AppPulse; 14: CA Mobile; 15: Apteligent; 16: Flurry; 17: AppsFlyer; 18 Yandex Metrica; 19 Adjust; 20: Ironsource; 21 Countly; 22: Sentry; 23: AndroidGodEye; 24: BlackCancary; 25: ArgusAPM
\end{tablenotes}
\end{threeparttable}
\label{tab:antipattern-apm}
\end{table*}

\noindent\textbf{Accessing sensitive data and files.}
Some APMs collect information from \code{logcat}. Even though the permission is deprecated since Android 4.1, it still works for legacy Android system versions prior to Android 4.0. In addition, the \code{logcat} contains the information from all running processes. Therefore, data from other processes is exposed to APMs and app developers, causing the leakage of users' privacy information.

We do not recommend APM vendors to access the \code{/proc/stat} file as it cannot be visited without the root permission since Android 8.0. APM vendors have other options (e.g., visit the file \code{/proc/cpuinfo}) for accessing the CPU information as presented in Sec. \ref{subsec:support-functions}. 
Although these approaches are not officially blocked by Android, using them can lead to privacy leaks. \cite{Zhou:2013}. 
For example, by accessing the file \code{/proc/cpuinfo}, we can obtain the device's CPU information, such as CPU family, CPU vendor id, and CPU model.

\subsection{Address Performance Anti-patterns in Android Apps.}\label{subsec:confront-anti-patterns}
Next, we explore whether APMs can assist developers in debugging and locating anti-patterns in apps. Anti-patterns are considered as bad programming practices in a program \cite{Erich:1993}. Inspecting whether APMs can resolve these anti-patterns helps APM vendors find room to improve their APMs.

We carefully select 8 performance anti-patterns of mobile apps from several empirical studies \cite{Liu:2014,Afjehei:2019,Chen:2018,Hecht:2015}. These anti-patterns are defined and confirmed by app developers and can be considered as the key issues concerned by developers in terms of app performance. We manually evaluate whether existing APMs can address them.

The list of anti-patterns is presented in Tab. \ref{tab:antipattern-apm}. Specifically, (1) \textit{GUI lagging} prevents user events from being handled in a timely way. This also triggers ANR errors; (2) \textit{energy leak} represents unexpected excessive consumption of battery power of an app; (3)  \textit{memory bloat} refers to the bug that can incur unnecessarily high memory consumption; (4) \textit{cyclic invocation} represents a frequently executed method in an invocation cycle; \textit{frequent invocation} refers to a method being frequently executed; (5) \textit{expensive callee}: a method is slow in executing its callees' code; (6) \textit{loading time} is the time for loading a resource or a UI display; (7) \textit{query a DB} stands for computing time spent for querying an item from a local database; and (8) \textit{UI overdraw}~\cite{Hecht:2015} represents catching the case that an app draws the same pixel more than once within a single frame.

We follow the following steps to determine whether an APM can detect a performance antipattern.

\noindent$\bullet$ STEP 1 (Learn from APM documentation): We first learn the performance anti-patterns from existing studies \cite{Trubiani:2018,Ahmed:2016,Willnecker:2015,Yao:2018} to understand their consequences. Then, we check whether the APM documentations clearly claim that they can detect these anti-patterns or they can detect the consequences caused by these performance anti-patterns. For example, the ``GUI lagging'' anti-pattern can result in ANR errors. Thus, APMs supporting ANR errors detection can detect this anti-pattern.

However, only using STEP 1 may introduce biases because sometimes an APM's documentation may not be consistent with its implementation. We perform additional checks in STEP 2 and 3.

\noindent$\bullet$ STEP 2 (Build samples): We build a sample app, which does not contain any performance anti-pattern. We term it ``\textit{base} app''. For each anti-pattern, we implement and add a module containing the anti-pattern to the \textit{base} app. We name the new app as \textit{anti$_{i}$}, where $i$ is the index of the anti-pattern (see Table \ref{tab:antipattern-apm}). For example, anti$_{1}$ represents the \textit{base} app with the GUI lagging;

\noindent$\bullet$ STEP 3 (Verification): To evaluate whether an APM can detect a performance anti-pattern $i$, we embed the APM to the \textit{base} app and the \textit{anti$_{i}$} app respectively. Then, we analyze whether the data collected by APMs suggests a performance anti-pattern. For example, if an APM finds that the \textit{anti$_{3}$} app consumes more memory than the \textit{base} app, it suggests a ``memory bloat'' anti-pattern.

The results are summarized in Table \ref{tab:antipattern-apm}. We can see that several common performance anti-patterns cannot be handled, including energy leak, time-consuming for database query, and UI overdraw. For ``energy leak'', the APMs under investigation cannot detect it as they don't collect battery usage data. Thus, we cannot rely on APMs to detect the energy leak antipattern. For ``query a local DB'', the existing APMs do not offer any DB related functions, such as the time needed for querying a database. Same for ``UI overflow'', if a pixel is drawn more than once, the state-of-art APMs do not provide any functions for recording it.

\begin{tcolorbox}[boxrule=1pt,boxsep=1pt,left=2pt,right=2pt,top=2pt,bottom=2pt]
Even though some anti-patterns can be well resolved by using APMs, there are still several common performance anti-patterns that cannot be addressed, including energy leak, time-consuming for database query, and UI overdraw. Therefore, there is room for APM vendors to improve their products.
\end{tcolorbox} 

\begin{figure*}[!htpb]
    \centering
    \includegraphics[width=0.8\textwidth]{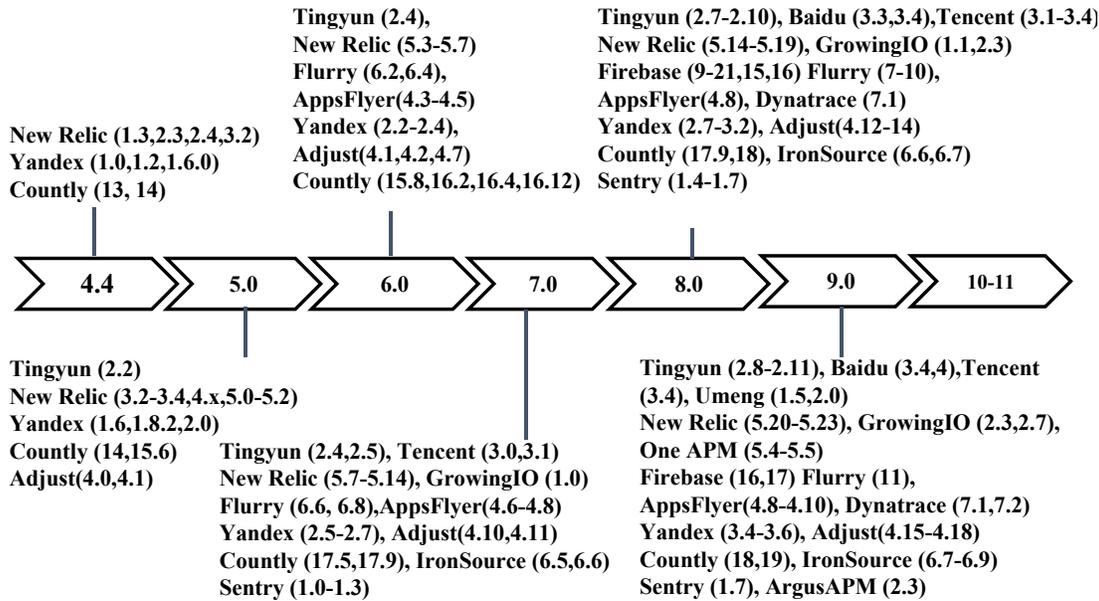}
    \caption{APM Evolution Timeline (map to Android version)}
    \label{fig:apm-evolution-process-timeline}
\end{figure*}

\subsection{Evolution of APMs} \label{subsec:apm-evolution}
Android itself frequently introduces new features and updates the existing system during its lifetime. With the updates of the Android system, new features are introduced and some APIs become deprecated. Therefore, it is worth investigating: (1) How do APMs evolve during their lifecycles? and (2) Do APMs evolve to respond to the changes on Android?

We collect all the available versions for the 25 APMs we studied. As APM SDKs are presented as \code{.jar} files, we leverage \code{PKGDiff} \cite{PKGDiff} to locate changes cross different versions. Then, we manually analyze the changes in the successive two versions and summarize them. 
To reason the changes and the intentions behind the changes, we analyze the changelogs and documentation of these APMs to ensure correctness.

By inspecting all the 25 APMs, we find there are 18 APMs\footnote{Other 7 APMs are OpenInstall, App Dynamcis, Site 24*7, AppPulse, CA, and Apteligent} whose historical versions are publicly available. Fig. \ref{fig:apm-evolution-process-timeline} shows the mapping relations between APM versions and Android versions. 
We can see that a single version of APM can spread across different Android versions.


After manually inspecting the changes upon APMs during their lifecycles, we learn that the changes fall into the following cases:

\noindent\textbf{1. Fixing the compatibility issues}: There are two sub-patterns for such compatibility-related changes: 

\underline{Pattern 1}. Fixing the compatibility issues with Android systems. When Android updates, the APMs may not be compatible with the latest version, which results in crashes at runtime. Thus, vendors update their APMs to fix such issues. 

\underline{Pattern 2}. Fixing the compatibility issues with other third-party libraries. Similar to other apps\cite{ZhanASE20}, APMs may use some third-party libraries. If they are not compatible with these libraries, APM vendors update their APMs to be compatible with the major third-party libraries.

\noindent\textbf{2. Additional features for additional application scenarios.} APM vendors update their products by supporting more scenarios. For example, APMs (e.g., Adjust, Flurry) add additional tracking APIs for developers to monitor the in-app purchase action. 

\noindent\textbf{3. Kotlin support and Native code support.} More and more apps are developed in the Kotlin language rather than Java. Thus, the APMs are evolved to support Kotlin-based apps. 

\noindent\textbf{4. Code optimization and bug fixing}: Another common practice is code optimization and bug fixing. As APMs are embedded into apps, APM vendors always attempt to limit the sizes of APMs and the resources required by APMs.

Besides, we also investigate whether the changes introduced by a new Android version break the functionalities of APMs and how APMs respond to these changes.

\noindent $\bullet$ Version $4.4 \rightarrow 5.0$: Since Android 5.0, the permission \code{GET\_TASKS} becomes deprecated. However, some APMs still request such permission from users.

\noindent $\bullet$ Version $7.0 \rightarrow 8.0$: Since Android 8.0, the file \code{/proc/stat} cannot be accessed without the root permission. However, APMs relying on this file to collect CPU usage do not make any adjustments.

\noindent $\bullet$ Version $9.0 \rightarrow 10-11$: Since Android 10.0, the sample rate allowed by API \code{getProcessMemoryInfo} is significantly limited. If the API is called faster than the limit, the same data as the previous call is returned. It suggests that APMs should adjust the invocation rate of the \code{getProcessMemoryInfo} API if used to collect memory usage. However, no APM makes the corresponding adjustments.

In summary, some version updates break APMs' functionalities, but APMs fail to make the corresponding changes to cope with such updates. As a result, these updates make some functions in APMs fail to work properly.

\begin{tcolorbox}[boxrule=1pt,boxsep=1pt,left=2pt,right=2pt,top=2pt,bottom=2pt]
By inspecting the evolution process of APMs, we summarize 5 key intuitions for APM vendors to update their APMs, including fixing comparability issues, hunting bugs, supporting Kotline, native code, supporting HTTP/2, optimizing code, and involving new features. Since some OS updates can break APMs' functionalities, APM vendors should carefully cope with the OS updates.
\end{tcolorbox} 

%% file: sections/methodolgy.tex
\section{Methodology}\label{sec:methodolgy}

\begin{figure}[!htpb]
    \centering
    \includegraphics[width=0.5\textwidth]{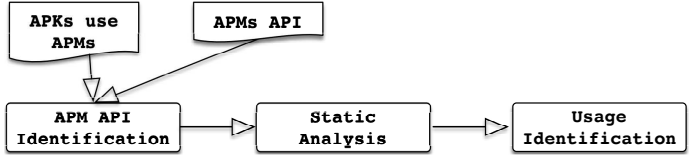}
    \caption{The Overview of APMHunter}
    \label{fig:apmhunter-overview}
\end{figure}

To understand how APMs are used in real apps, we develop \code{APMHunter}, an automated tool, to detect the usages of APMs in apps. The overview of APMHunter is shown in 
Fig. \ref{fig:apmhunter-overview}. APMHunter contains three major components: \code{APM API Identification}, \code{Static Analysis}, and \code{Usage Identification}.

\noindent$\bullet$ The \code{APM API identification} module aims at detecting whether any APM is used in an app;

\noindent$\bullet$ The \code{static analysis} module is a general framework to analyze a given app that uses APMs; and

\noindent$\bullet$ The \code{usage identification} module records the usage patterns of APMs and checks misuse of APMs (i.e., collect sensitive data with APMs).

\subsection{APM API Identification}\label{subsec:api_identification}

\noindent\textbf{Motivation.} This component aims at identifying APMs used in apps, which helps us spot apps that leverage APMs. Since code obfuscation techniques transform code into an obscure format \cite{Zhang:2014,Linares-Vasquez:2014}, they are widely used by commercial apps. To analyze obfuscated apps, we design the following obfuscation-resilient approach. If an app is packed, we will first use existing tools \cite{PackerGrind20,xue2017adaptive,DexHunter} to unpack it.

\noindent\textbf{Intuition.} As reported in \cite{Backes:2016,Zuo:2019}, there are usually at least two invariants that are preserved during obfuscations.

\noindent$\bullet$ The first invariant is the class inheritance relation. Although the names of classes, methods, parameters, and variables can be obfuscated, class inheritance relations remain stable even with obfuscation. For example, if class $A$ is a superclass of class $B$, after obfuscation, $A$ is still the superclass of $B$; and 

\noindent$\bullet$ The second invariant is the caller-callee relation. Inside a program, the invocation from a caller to its callee does not change. For instance, a method \code{downloadVideo} calls two methods \code{Util.isNetworkAvailable} and \code{StringBuilder.append}, the caller-callee relations from \code{downloadVideo} to \code{Util.isNetworkAvailable} and \code{StringBuilder.append} donot change during obfuscation.

\noindent\textbf{Methodology.} We perform the following steps to identify APM usages in apps:

\noindent$\bullet $ \textbf{Step 1.} We first build the signatures of all methods in all 25 APMs. It results in a list of signatures for the APIs in these APMs. The way to generate the signature for a given method is presented in Alg. \ref{alg:method-signature-generation};

\noindent$\bullet $ \textbf{Step 2.} For a given app, we build the signatures of all methods in the app;

\noindent$\bullet $ \textbf{Step 3.} Last, we leverage the signature lists of APMs (results in \textbf{Step 1}) as the search queries for detecting the usage of APMs in an app.

Our signature construction for a single method is presented in Alg. \ref{alg:method-signature-generation}. As mentioned in \textbf{Step 1} and \textbf{Step 2}, Alg. \ref{alg:method-signature-generation} is used to generate signatures for methods in APMs and apps. Note that, the terms ``class'' and ``type'' are used interchangeably as a type in Java is represented by a class. We use ``class'' to represent both classes and interfaces rather than distinguishing them.

\begin{algorithm}[!htpb]

\SetKwFunction{cumprod}{cumprod}
\SetKwFunction{length}{length}
\SetKwFunction{zeros}{zeros}
\SetKwFunction{ceil}{ceil}

\SetKwInOut{Input}{Input}
\SetKwInOut{Output}{Output}

\caption{GETFUNCSIG}
\label{alg:method-signature-generation}
\Input{\xvbox{2mm}{$f_{0}$} -- target function; $\mathcal{C}_{s}$ -- system classes; $\mathcal{F}_{s}$ --  system functions;
}

  \BlankLine 
  
  fSigBuf $\leftarrow$ $\emptyset$  \tcp{fSigBuf is a temporary String buffer}

  $\mathcal{L} \leftarrow \emptyset$  \tcp{$\mathcal{L}$ is a temporary variable}
  
  $t_{0}$ $\leftarrow$ GETHOSTCLASSTYPE $\left ( f_{0} \right )$
  
  fSigBuf $\leftarrow$ fSigBuf $\cup$ TYPEENCODING $\left ( t_{0}, t_{0}, \mathcal{C}_{s}\right)$
  
  \tcc{Get and encoding parameter}
  \For{$t_{p} \leftarrow $ {GETPARAMETERTYPE} $\left( f_{0}\right)$ }{
     fSigBuf $\leftarrow$ fSigBuf $\cup$ TYPEENCODING $\left ( t_{0}, t_{p}, \mathcal{C}_{s}\right)$
  }
  
  \tcc{Get and encoding return}
  $t_{r}$ $\leftarrow$ GETRETURNTYPE $\left( f_{0} \right)$

  fSigBuf $\leftarrow$ fSigBuf $\cup$ TYPEENCODING $\left ( t_{0}, t_{r}, \mathcal{C}_{s}\right)$
  
  \tcc{Get and encoding all callees in the method}
  \For{each callee $f_{i}$ $\in$ GETCALLEE $\left( f_{0} \right)$}{
    \If{$f_{i}$ $\in$ $\mathcal{F}_{s}$}{
        $\mathcal{L} \leftarrow$  $\mathcal{L} \cup$ name ($f_{i}, argType(f_{i}), retType(f_{i})$) 
    }
    \Else{
        $\mathcal{L} \leftarrow$ $\mathcal{L} \cup$  GETFUNCSIG $\left(f_{i}, \mathcal{C}_{s}, \mathcal{F}_{s}\right)$
    }
  }
  
  fSigBuf $\leftarrow$ fSigBuf $\cup$ SORTNEXCLUDERE$\left(\mathcal{L} \right)$  \tcp{SORTNEXCLUDERE: sort and remove redundant signatures}

  \Return fSigBuf
\end{algorithm}

\noindent$\bullet$ \textbf{Input}: Our algorithm requires three inputs: $\xvar{f}$ represents the method for generating signature; $\mathcal{C}_{s}$ gives the all system classes; and $\mathcal{F}_{s}$ denotes all system functions;

\noindent$\bullet$ \textbf{Line 1-4}: We first get the type ($t_{0}$) of $f_{0}$'s host class (i.e., GETHOSTCLASSTYPE in Line 3). Next, we invoke the TYPEENCODING method (see Alg.\ref{alg:type-encoding}) to encode the type $t_{0}$;

\noindent$\bullet$ \textbf{Line 5-7}: Then, we encode the parameters that declared in the method $f_{0}$ with the TYPEENCODING method;

\noindent$\bullet$ \textbf{Line 8-9}: We embed the \code{return} information into $f_{0}$'s signature;

\noindent$\bullet$ \textbf{Line 10-17}: If method $f_{0}$ invokes any methods, we iteratively collect the signatures of these callees. Then, we add these signatures to the temporary variable $\mathcal{L}$. However, if a callee is a system function (i.e., Android system API), we directly add the method signature to the temporary variable $\mathcal{L}$; and

\noindent$\bullet$ \textbf{Line 18-19}: Then, we sort and remove redundant signatures in $\mathcal{L}$ and append the results to fSigBuf. We sort and remove redundant signatures in $\mathcal{L}$ as: (1) we only consider the method invoked in $f_{0}$ regardless of their order; and (2) if a method is invoked multiple times, we only count once. As our task in this module is to identify certain methods (i.e. APM APIs) that are invoked in apps, we only need to consider each signature once.  

 
\begin{algorithm}[!htpb]

\SetKwFunction{cumprod}{cumprod}
\SetKwFunction{length}{length}
\SetKwFunction{zeros}{zeros}
\SetKwFunction{ceil}{ceil}

\SetKwInOut{Input}{Input}
\SetKwInOut{Output}{Output}

\caption{TYPEENCODING}
\label{alg:type-encoding}
\Input{\xvbox{2mm}{$c_{h}$} -- class; $c_{t}$ -- target class; $\mathcal{C}_{s}$ -- classes defined in app;
}

  \BlankLine 
  
  tBuf $\leftarrow$ $\emptyset$  \tcp{tBuf is a temporary String buffer}

  $\mathcal{L} \leftarrow \emptyset$  \tcp{$\mathcal{L}$ is a temporary String buffer}
  
  \If{$c_{t} \in$ $\mathcal{C}_{s}$}{
     tBuf $\leftarrow$ tBuf $\cup$ name($c_{t}$)
  }
  \Else{
     $c_{p}\leftarrow$ GETSUPERCLASS ($c_{t}$)
     
     tBuf $\leftarrow$ tBuf $\cup$ TYPEENCODING ($c_{h},c_{p},\mathcal{C}_{s}$)
     
     \For{$c_{i}\leftarrow$ GETINTERFACES ($c_{t}$)}{
        $\mathcal{L} \leftarrow$ $\mathcal{L}\cup$ TYPEENCODING ($c_{h},c_{i},\mathcal{C}_{s}$)
     }
     tBuf $\leftarrow$ tBuf $\cup$ SORT($\mathcal{L}$)
  }
  
  \Return tBuf
\end{algorithm}

The function TYPEENCODING (Alg. \ref{alg:type-encoding}) is designed for encoding any type defined in the app. 

\noindent$\bullet$ \textbf{Input}: The TYPEENCODING's input consists of the host class $c_{h}$, which represents the current class context of the $c_{t}$ (i.e., we currently encode the class $c_{t}$ which is used in class $c_{h}$), target class $c_{t}$, which represents the type under encoding; and  $\mathcal{C}_{s}$, which contains all classes defined in the app;

\noindent$\bullet$ \textbf{Line 1-5}: If the type (i.e. $c_{t}$) to be encoded is a system type (i.e. Android system API), we return the name of the type;

\noindent$\bullet$ \textbf{Line 6-13}: If the type (i.e. $c_{t}$) to be encoded is defined in the app, we encode its superclass $c_{p}$ (if exists), and append the encoding of $c_{p}$ to the temporary result tBuf. If $c_{t}$ inherits any interface, we also encode all interfaces, and append the encoding results to tBuf. We sort and remove redundant signatures in $\mathcal{L}$ as we only consider the interfaces inherited by $c_{t}$ regardless of their order; and

\noindent$\bullet$ \textbf{Line 14}: The temporary variable tBuf is returned as the coding of the type $c_{t}$.

When building method signatures with Alg. \ref{alg:method-signature-generation}, we replace non-system identifiers/names with symbol 'x' in signatures to reduce the side-effects introduced by obfuscation. For example, class name \code{com.networkbench} is replaced by \code{X.X}.

After encoding the methods in an app and an APM with Alg. \ref{alg:method-signature-generation}, we search the initialization method of an APM inside the app. For example, the UMeng APM can be initialized by invoking the \code{UMConfigure.init()} API. If method $m$ in app invokes the \code{UMConfigure.init()}, method \code{UMConfigure.init()}'s signature must appear in $m$'s signature as \code{UMConfigure.init()} is a callee of $m$ (see Alg. \ref{alg:method-signature-generation} (Line 10-17)). As a result, we know that the APM is used in the app.

\subsection{Static Analysis}\label{subsec:static_analysis} 
\noindent\textbf{Motivation.} After confirming an app contains an APM, the next step is to characterize the APM usages in the app. We intend to collect the following information: (1) the position where APMs are initialized; (2) the leak of users' private data thought the APM; and (3) the APM APIs used in the app (See \textsection \ref{subsec:usage-identification}). To achieve this goal, we conduct a static analysis on the app to build its inter-procedural control-flow graph (ICFG).

\noindent\textbf{Methodology.} To build the ICFG and conduct the data flow analysis on an app, we perform the following steps: (1) locating entry points for the app; (2) performing static analysis; and (3) exploring inter-component communications in the app.

\subsubsection{Entry Points}\label{subsec:entry-point-discovery}
To build the ICFG for an app, we first resolve its entry points. Unlike desktop Java applications, Android apps do not have explicit entry points (e.g., \code{main} method). 
The entry points of an app are derived from two aspects \cite{Holavanalli:2013}: (1) lifecycle methods in Android components (i.e., \code{Activity}); and (2) user interface (UI) events handler callbacks. Specifically, we leverage the state-of-art tool EdgeMiner \cite{Cao:2015} to explore these two types of entry points.

\subsubsection{Build the ICFG and Perform Data-flow Analysis}\label{subsec:icfg-analysis}
The next step is to build the ICFG for an given app. The ICFG contains control flow information of the given app. Besides, we perform the data flow analysis on the app, and append the data flows to the ICFG. This process is implemented based on the FlowDroid~\cite{Arzt:2014}, which is a widely used Android app analysis framework based on Soot (Java optimization framework) \cite{Soot}.

\subsubsection{Inter-component Communication (ICC)}\label{subsec:intent-resolve}
Inside an app, data can be sent cross different components through \code{Intents}. Since \code{Intents} in an app introduce additional data flows to the ICFG, we append data flows introduced by \code{Intents} to our ICFG. An \code{Intent} can be \textit{explicit} or \textit{implicit}. In an \textit{explicit} \code{Intent}, the target is given by an explicit class name. \code{APMHunter} obtains the class name carried by the Intent and then links the target class with the Intent. An \textit{implicit} \code{Intent} only specifies the functionality that it wants to invoke instead of the class name of the target component. To infer the target of an \textit{implicit} Intent, we adopts the \code{IC3} \cite{Octeau:2013} tool. \code{IC3} transforms the ICC problem into a Multi-Valued Composite (MVC) constant propagation problem (i.e., finding all possible values of objects concerned at a certain program point). 

IC3 resolves the MVC constant propagating problem with the COnstant propAgation Language (COAL) and then employs a COAL solver to solve the problem. By building a Program Dependence Graph (PDG) and performing an MVC data flow analysis, IC3 can infer the arguments in an implicit Intent, and then find the target component for the Intent.
We first run IC3 to collect the inter-component communications between components. The results can be bound with FlowDroid by leveraging the API \code{IC3ResultLoader}. Once this setup is accomplished, the data flows introduced by ICC can be appended to our ICFG.

\subsubsection{String Analysis}\label{subsec:string-analysis}
\noindent\textbf{Motivation.} As introduced in \textsection\ref{sec:apm}, developers can use APMs to collect runtime values with the built-in logging functions. Hence, we conduct the string analysis to explore the data collected by developers. Such data can reflect developers' intention of using APMs and determine the correlation between APMs and apps (e.g., category, functionality).

\noindent\textbf{Methodology.} To capture the values collected in the app, we leverage Violist \cite{Li:2015}, a String analysis framework for Android, to perform String analysis. Specifically, Violist separates the \textit{representation} and \textit{interpretation} of string operations. To compute the value of a string, Violist defines an Intermediate Representations (IRs) to capture string operation. After computing the string value in the IR format, the result (in IR format) is translated to a string. 
Finally, the String analysis function (from Violist) is integrated into our APMHunter framework.





\subsection{Usage Identification}\label{subsec:usage-identification}

\noindent\textbf{Motivation.} The key goal of the \textit{usage identification module} is to understand how APMs are used by developers in terms of the APIs used by developers and their context information (e.g., where the APIs are instrumented).

\noindent\textbf{Methodology.} 
APMHunter records the following usages of an APM for a given app:

\textbf{(I) The position of an APM's initialization.}
When an APM is initialized by an app, it starts monitoring the performance of the app. Therefore, locating an APM's initialization can provide insights on developers' usage patterns of an APM and help us detect potential APM misuse.

\textbf{(II) The privacy leaks through APMs.}
We detect the privacy leaks from two aspects: 
(1) privacy leaks from user inputs. For example, if there is a field for a password, the data of the field can be leaked to the APM in the app. Consequently, developers receive sensitive data from users; and
(2) privacy leaks from permission-related APIs. For example, the location data can be obtained with method \code{getLastknowLocation()}\cite{Android}, which can be leaked to developers via APMs. Such leakage is defined as permission-related privacy leak as the method \code{getLastknowLocation()} is associated with the ACCESS\_COARSE\_LOCATION or ACCESS\_FINE\_LOCATION permission.

Next, we illustrate the process of these two types of privacy leakage detection.

\begin{figure}[!htpb]
    \centering
    \includegraphics[width=0.36\textwidth]{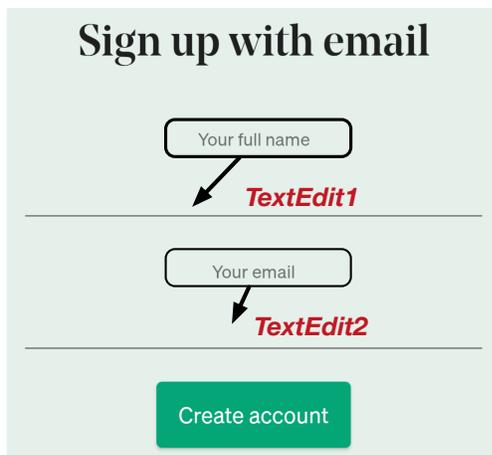}
    \caption{Adjacent Label sample from Medium App}
    \label{fig:adjacent-hint-android}
    \vspace{-1ex}
\end{figure}

\noindent$\bullet$ Users' account information and profiles. It reveals users' personal information in an app, which includes username, first name, last name, password, email account, birthday and phone number;

\noindent$\bullet$ Location information. It represents users' home addresses. Different from the system derived location (i.e., latitude and longitude), here we focus on the delivery addresses/home addresses, which are input by users; and

\noindent$\bullet$ Users' financial information. The financial information is mainly related to users' payment information, such as credit card numbers, expiration date, and security code.

To detect the sensitive information, we follow the idea in UIPicker \cite{UIPicker} to perform the GUI analysis\cite{ZhouASE20}. First, for a given app, we parse the resources file (e.g., layout files, strings) in the app. Next, we identify the textual semantics of UI element with natural language processing and fetch corresponding private data. Specifically, a Support Vector Machine (SVM) classifier is built with the supervised machine learning. It takes a UI element's context in the whole layout into consideration to determine whether an element is privacy-sensitive. If a UI element is associated with the private data, we track the access of the UI element starting from the \textit{findView} APIs, including \code{findViewById()}, \code{findViewWithTag()}, \code{findViewsWithText()}, by adding these to \code{SourceAndSink} file as source. The \code{SourceAndSink} is served as the input for FlowDroid module in APMHunter to detect privacy leaks. The evaluation of sensitive UI element identification is presented in Sec. \ref{subsec:evaluation-apmhunter}.

\noindent\textbf{Privacy leaks from permission-required data.} We use the static analysis to identify the permission-required sensitive data leaked to APMs. PScout \cite{Au:2012}, a widely used API-Permission mapping set, is leveraged to collect system APIs whose executions require certain permissions to be granted. For example, method \code{getLastKnowLocation()} requires the ACCESS\_COARSE\_LOCATION or ACCESS\_FINE\_LOCATION permissions to be granted. Then, \code{APMHunter} builds the \code{SourcesAndSinks} file for finding privacy leak paths using FlowDroid \cite{Arzt:2014}.

Next, we use an example to illustrate why settings can detect the privacy leak. If an app leverages \code{getLastKnowLocation()} to obtain the users' location, then the location data is passed to an APM. As a result, users' privacy is leaked. Such privacy leakage is represented by a valid path from \code{getLastKnowLocation()} to an APM logging API, which can be detected by our tool \code{APMHunter}.

\textbf{(III) Characterizing the used APM APIs and their context information.} 
To infer the developers' intentions of using APMs, we record the APM APIs used in apps. For example, the API \code{setRevenue} in Adjust APM is used to track the revenue for an app. By recording the usage of this API, we learn that the developers aim at tracking their revenues. 

Specifically, we record the following information for understanding the APM usages:

\noindent$\bullet $ The methods and classes/interfaces that use APM APIs. For methods, we check whether these methods are lifecycle methods or inheritance methods of lifecycle methods in Android \cite{Android}. For classes/interfaces, we check whether these classes represent app components or app instances;

\noindent$\bullet $ We record the code segments under monitoring. For instance, sometimes, textcolor{red}{developers leverage the tracking APIs in APMs to understand how app users interact with their apps, and which parts in apps draw users' interests; and}

\noindent$\bullet $ We also infer the variables collected by developers via APMs using string analysis. By inspecting variables collected by developers, we can infer developers' intentions.

\subsection{Implementation}
APMHuner is built atop FlowDroid \cite{Arzt:2014}. In APMHuner, we link the entry point methods in an app with the \code{dummyMain}. The \code{dummyMain} method, a faked main method provided by FlowDroid, allows developers to traverse the ICFG through the \code{dummyMain} rather than starting from all entry points. FlowDroid also provides an interface (see class \code{Ic3ResultLoader}) for loading the results produced by IC3. The string analysis module (from Violist) and additional features for exploring APM usages are integrated into our tool APMHunter. More implementation details of APMHunter can be found on our project page.


\subsection{Evaluation on APMHunter}\label{subsec:evaluation-apmhunter}
\subsubsection{Accuracy on APM detection}\label{subsubsec:accuracy-apm-detection}
We evaluate the accuracy of APMHunter by using 500 randomly-selected apps from Google Play. 
\noindent To determine whether an app is obfuscated or not, we disassemble each app with Apktool \cite{Apktool}, and then manually check whether its package name is obfuscated or not. As shown in Table \ref{tab:sample-app-benchmark}, there are 123 obfuscated apps and 377 non-obfuscated apps. Note that, since identifying the obfuscated apps is non-trivial process and difficult. For obfuscated apps here, we just consider the apps with package name obfuscation.

\begin{table}[!htpb]
\centering
\caption{The Benchmark of Selected Apps}
\begin{tabular}{|c|c|c|c|}
\hline
\textbf{}               & \textbf{With APM} & \textbf{Without APM} & \textbf{Total} \\ \hline
\textbf{Obfuscated Apps}     & 70               & 53                   &  123       \\ \hline
\textbf{Non-obfuscated Apps} & 122               & 255                  &    377         \\ \hline
\textbf{Total}     & 192                & 308                   & 500            \\ \hline
\end{tabular}
\label{tab:sample-app-benchmark}
\end{table}


\begin{figure}[!htpb]
    \centering
    \includegraphics[width=0.5\textwidth]{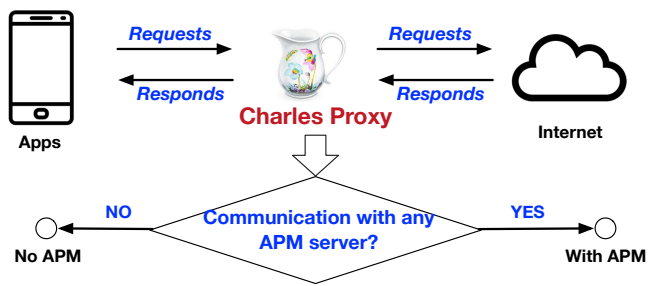}
    \caption{Workflow for Building the Benchmark}
    \label{fig:apm-usage-detection-charles}
\end{figure}

Fig. \ref{fig:apm-usage-detection-charles} shows the workflow for building the benchmark, which is used for evaluating the performance of APMHunter. First, we set up a CharlesProxy \cite{Charles} and connect the target device to the CharlesProxy. CharlesProxy is a cross-platform HTTP debugging proxy server application, which can capture the communication between target device and remote servers. Once an app sends requests to a remote server or receives responses from a remote server, the CharlesProxy captures these packets. If an app uses a specific APM, the APM sends the collected data to the APM's remote server. The captured packets sent from an app to a specific APM server suggest the APM used by the app. For example, if an app contains the Apteligent APM, we can capture the packets sent to \code{appload.ingest.crittercism.com} via CharlesProxy. Note that Apteligent was named as crittercism. In this way, we can determine whether an app uses an APM even if the app is obfuscated. In practice, each app will be run automatically for one hour with our python script that is built upon the Android Monkey framework \cite{AndroidMonkey}. We then analyze all packets captured by CharlesProxy to determine whether an app uses certain APMs. As APMs periodically upload data collected to remote servers, the execution coverage of the target app does not affect the result.

\begin{table}[!htpb]
\centering
\caption{The Performance of APMHunter on APM Identification}
\begin{threeparttable}
\scalebox{0.8}{
\begin{tabular}{|c|c|c|c|}
\hline
\textbf{APMHunter}                              & \textbf{With APM (Benchmark)} & \textbf{Without APM (Benchmark)} & \textbf{Total} \\ \hline
OA-With                & 60                       & 4                           & 64             \\ \hline
NO-With   & 108                       & 0                        & 108          \\ \hline
OA-Without          & 10                        & 49                      & 59            \\ \hline
NO-Without & 14                        & 255                        & 269      \\ \hline
\textbf{Total}                         & 192                           & 308                             & 500            \\ \hline
\end{tabular}
}
\begin{tablenotes}
    \item OA-With: obfuscated apps that use APMs; \item NO-With: None-obfuscated that apps use APMs;
    \item OA-Without: obfuscated apps that do not use any APM;
    \item NO-Without: None-obfuscated app that do not use any APMs;
    
\end{tablenotes}
\end{threeparttable}
\label{tab:accurarcy-apm-identification}
\end{table}



By inspecting the packets sent from apps, we find that 192 apps use APMs, and 308 apps do not employ any APM as shown in Table \ref{tab:sample-app-benchmark}. We apply APMHunter to these sample apps in order to evaluate its performance. The results are shown in Table \ref{tab:accurarcy-apm-identification}. In Table \ref{tab:accurarcy-apm-identification}, the columns represent the statistical results by referencing the benchmark, the rows represent the statistical results given by the APMHunter. For example, 60 (row 1, column 1) represents that there are 60 obfuscated apps that use APMs, and APMHunter correctly identifies them. 10 (row 3, column 1) represents that there are 10 obfuscated apps that use APMs, but APMHunter fails to find them. The precision of APMHunter is 97.7\% (168/172) and the recall of APMHunter is 87.5\% (168/192).

%

After manually inspecting 28 apps that APMHunter fails to make the correct decisions, we have the following observations. The reason why APMHunter cannot reach 100\% precision is due to the use of other libraries (e.g., Google GMS). When an app uses a third-party library, especially Google GMS or Google Ads, the Firebase APM is invoked by the third-party library. However, the original app does not use the Firebase APM. 
For non-obfuscated apps, APMHunter can distinguish that the APM is invoked by a third-party library. However, APMHunter cannot determine whether the APM is invoked in the third-party library or the host app if the app is obfuscated, and thus it may make incorrect decisions.

The encoding algorithm in the APM identification module is version insensitive. That is, the method signature of a method in an APM's two different versions can be different. Our encoding algorithm is built upon two invariants: class inheritance relations and caller-callee relations, which can be changed during evolution. If an app adopts an out-of-date APM, APMHunter may not identify the APM in use. This is the obstacle in achieving a higher recall. The reason why we do not support all versions of APMs is that some vendors only provide the latest SDK versions and the previous versions are no available.

%

\begin{table}[!htpb]
\caption{The Performance of APMHunter on Detecting Sensitive UI Elements}
\scalebox{0.8}{
\begin{tabular}{|c|c|c|}
\hline
\textbf{}                        & \textbf{Sensitive UI element} & \textbf{Insensitive UI element} \\ \hline
\textbf{Sensitive UI (APMHunter)} & 383              & 16                 \\ \hline
\textbf{Insensitive UI (APMHunter)}      & 24                & 149                   \\ \hline
\end{tabular}
}
\label{tab:apmhunter-performance-sensitive-ui}
\end{table}

\subsubsection{Accuracy of identifying sensitive UI elements}\label{subsubsec:accuracy-sensitiveui-detection}
We use the 500 randomly selected apps to evaluate whether sensitive UI elements can be captured by our tool. We install these apps on a device and inspect the UI elements in apps. Among all samples, there are 129 apps whose UIs are not in English. By manually validating the rest 371 apps, we find 407 sensitive UI elements. Even though not all 371 apps contain APMs, we still consider them in this experiment, because the target of this experiment is assessing whether our tool can correctly identify sensitive UI elements from apps. The performance of our tool on detecting sensitive UI elements is displayed in Tab. \ref{tab:apmhunter-performance-sensitive-ui}. The precision is 95.9\% (383/399) and the recall is 94.1\% (383/407).

The false positive rate is 4.1\%. We manually inspect the 16 elements that APMHunter fails to make the correct decision and find the reason: APMHunter cannot capture the semantic context of the app, which misleads the tool to make the incorrect decision. For example, in a travel app, the ``phone number'' field refers to the phone number of a hotel rather than a user's phone number. Moreover, in some game apps, the ``name'' fields can refer to virtual names or other meaningless strings. For example, there is a name combiner app, which asks users to input two names and then combines them into one. In this app, the name field can be any string rather than a real user name.

The false negative rate is 5.9\%. We find that 24 UI-sensitive elements cannot be identified by our tool and find the reasons as follow: (1) some very low-frequency texts or abbreviations cannot be inferred by the privacy-related analysis, such as DLNO (i.e., Drive License Number), SSN (Social Security Number, which is used in US); and (2) some apps use icons to provide semantic information for the required fields. For example, a mail icon indicates that uses have to input their emails in the EditText field.

\subsubsection{Accuracy of privacy leak analysis}
\noindent As presented in Sec. \ref{subsubsec:accuracy-apm-detection}, 192 apps (out of 500 sample apps) use APMs. Among them, there are 141 whose UIs are in English. Therefore, our analysis is based on these 141 apps. 

\begin{figure}[!htpb]
    \centering
    \includegraphics[width=0.5\textwidth]{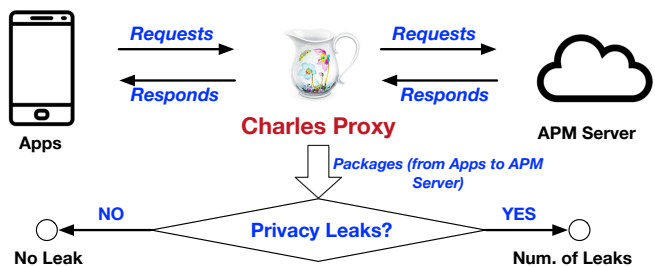}
    \caption{Workflow for evaluating the accuracy of APMHunter in terms of privacy leaks detection.}
    \label{fig:apm-privacy-leaks-detection}
\end{figure}

\noindent Fig. \ref{fig:apm-privacy-leaks-detection} shows the workflow for evaluating the accuracy of APMHunter in terms of privacy leaks detection. We use CharlesProxy to collect packets from apps to APM servers. By inspecting the packets collected, we can determine whether users' private data is leaked to APM servers. Specifically, we first build a virtual profile for testing, which contains the following fields: email, name (both first and last name), sex, age, race, phone number, ID, home address, credit card number, credit card CVV number, location. For a given app, we connect the Pixel phone to ChralesProxy and install the app. Next, we run the app and input the data from the virtual profile when required. For example, if the app needs us to register an account, we use the data in the virtual profile to build the account. As aforementioned, the CharlesProxy can listen to the communication between apps and APM servers. We collect the packets sent from apps to their APM servers. If sensitive data (i.e., data from virtual profile) is found in the packets captured by CharlesProxy, we consider it to be a privacy leak. 

\noindent By verifying each app, we find 91 leaks from 16 apps. APMHunter correctly reports 77 leaks out of 91 leaks. The recall of APMHunter is 84.6\% in terms of detecting privacy leaks. We find the following reasons lead to this: (1) third-party UI framework: some apps leverage third-party UI frameworks (e.g., Butter Knife) rather than the default UI frameworks. As APMHunter does not support third-party UI frameworks at this stage, some sensitive data leaked in this way cannot be detected with APMHunter; (2) some apps use a medium (e.g., a file, SharedPreference) to transfer data, which cannot be captured. For example, sensitive data is first written to a file and then the data is retrieved from the file. The precision of APMHunter is 96.2\% as we find that 3 leaks reported by APMHunter are not real leaks. This is because FlowDroid, upon which APMHunter is built. Thus, it cannot handle array indices precisely. 



%% file: sections/empiricalstudy.tex
\section{Empirical Study}\label{sec:empiricalstudy}

We guide our empirical study by answering five research questions (RQs), which are organized by three aspects:

\noindent$\bullet$ \textbf{1}: The popularity of APMs in the wild (RQ 1);

\noindent$\bullet$ \textbf{2}: How APMs are used in practice. Specifically, we are interest in: comparing logging frameworks with APMs (RQ 2); discussing consequences of using multiple APMs (RQ 3); and learning the privacy issues raised by APMs (RQ 4); and 

\noindent$\bullet$ \textbf{3}: What is the performance overhead of using APMs (RQ 5).

\subsection{Popularity of APMs in the Wild}\label{subsec:popularity}
\noindent\textbf{$\bigstar $ RQ 1. How prevalent are APMs in Android apps?}

\noindent\textbf{Motivation.}
We are interested in understanding whether APMs have been widely adopted in Android apps by answering the following sub-questions:

\noindent$\bullet $(\textbf{1-A}): 
Are APMs widely adopted by Android apps? 

\noindent$\bullet $(\textbf{1-B}): How APMs are ranked according to popularity? 

\noindent\textbf{Methodology.} (\textbf{1-A}) We implement our obfuscation-resilient approach for APM detection (see \textsection\ref{sec:methodolgy}) in APMHunter and use it to characterize the APM usage in a large set of apps. (\textbf{1-B}) Then, we calculate the popularity of APMs based on the results from the previous step. 

\begin{figure}[!htpb]
    \centering
    \includegraphics[width=0.5\textwidth]{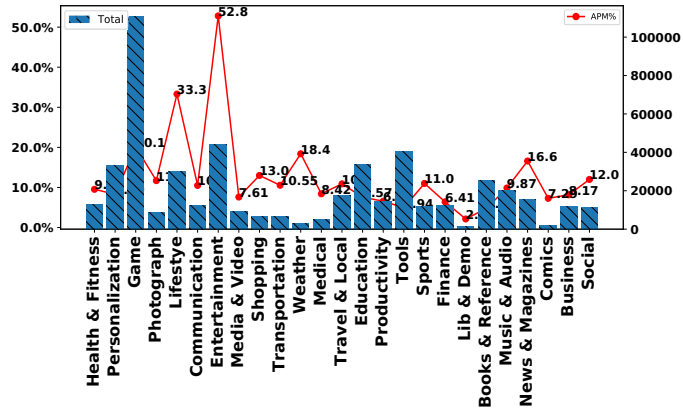}
    \caption{APM Usage across App Categories.}
    \label{fig:apm-usage-by-category}
\end{figure}

\noindent\textbf{Subject Apps.} We randomly crawl 500,000 Android apps from Google Play, which cover 25 main app categories. Note that, in Google Play, the game apps are grouped in different sub-categories. In our taxonomy, we consider all the sub-categories of game as one (i.e., Game). The sizes of these apps range from 100KB to 1.2 GB. The distribution of apps across the 25 categories is shown in Fig.~\ref{fig:apm-usage-by-category}.

\noindent\textbf{Results.} (\textbf{1-A}) From the 500,000 apps, we find that 55,722 apps (11.1\%) use APMs. The categories of these apps are summarized in Fig. \ref{fig:apm-usage-by-category}. We observe that APMs are most popular in the apps from the following five categories: \textit{Entertainment} (52.8\%), \textit{Lifestyle} (33.3\%), \textit{Game} (20.1\%), \textit{Weather} (18.4\%) and \textit{News \& Magazines} (16.6\%). On the contrary, APMs are infrequently used in the apps from the following categories: \textit{Library \& Demos} (2.13\%), \textit{Book \& Reference} (4.76\%), \textit{Finance} (6.41\%), \textit{Productivity} (6.57\%), and \textit{Comics} (7.29\%). 

As developers mainly use APMs to monitor the performance of their apps, we find some clues about the different adoption rates:

\noindent$\bullet$ \textbf{User Experience}: For Entertainment and Game apps, the runtime performance of these apps can strongly influence user experience. Poor user experience decreases satisfaction, loyalty, and credibility. They can leverage APMs to detect performance bottlenecks and fix bugs promptly. Thus, leveraging APMs can assist developers in improving user experience.

\noindent$\bullet$ \textbf{Performance}: For Weather, Lifestyle, News apps, they request data (e.g, weather data, news data) from remote servers and display them to end-users. Thus, monitoring and ensuring the network performance would be the major concern for developers to adopt APMs.

For the most time, apps from Library, Demos, and Book categories run locally. The runtime performance may not be as important as apps from other categories. As a result, the adoption rate for using APMs is low.

\begin{figure}[!htpb]
    \centering
    \includegraphics[width=0.45\textwidth]{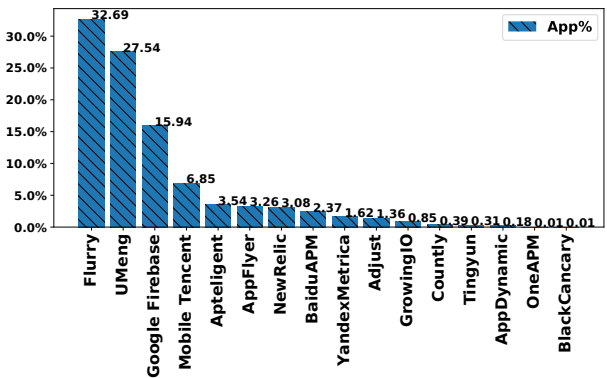}
    \caption{Share of APMs in terms of Popularity}
    \label{fig:apm-popularity}
\end{figure}

\noindent(\textbf{1-B}) For all the 25 APMs we studied based on 224,039 app samples, we calculate the frequency of each APM. The share of each APM is displayed in Fig. \ref{fig:apm-popularity}. The top popular APMs are: Flurry APM (32.69\%), UMeng (27.54\%), Google Firebase (15.94\%), Mobile Tencent (6.85\%), Apteligent (3.54\%), and AppFlyer (3.26\%). Furthermore, some APMs are nearly not adopted (<0.01\%) in the sample apps, including AndroidGodEye, AppPulse, ArguAPM, CA Mobile, Dynatrace, Ironsource, OpenInstall, Sentry, Site 24*7.

\noindent$\bullet$ \textbf{Commercial APMs. vs. Open Source APMs.} Besides, we compare the use of commercial APMs and open-source APMs. Surprisingly, we find that developers overwhelmingly choose commercial APMs comparing to open source APMs, i.e., 99.6\% of the apps are with commercial APMs, whereas only 0.4\% of all apps choose open source APMs. Based on the comparison between open-source and commercial APMs, we can conclude the following reasons for choosing commercial APMs:

\noindent$\bullet$ \textbf{Functionality.} We find that in most cases commercial APMs contain all functions in open-source APMs. It means open-source APMs do not have any unique part to make them competitive;

\noindent$\bullet$ \textbf{Usability.} Most open-source APMs do not provide any dashboard for collecting and analyzing data. Developers have to deploy or even build the dashboard on their own, which can be time-consuming. However, for commercial APMs, a dashboard is provided where all collected data can be displayed; 

\noindent$\bullet$ \textbf{Costs.} Most commercial APMs allow developers to deploy one app for free. Developers are charged if they have more than one app published. However, this can be easily bypassed by creating multiple accounts with different email accounts; and

\noindent$\bullet$ \textbf{Technical Support.} Besides, technical support is another factor that can affect the choice of APMs. As open-source APMs lack such support, developers have to maintain the APMs (e.g., bug fixing) on their own.

Therefore, we can conclude that compared to the service from commercial APMs, current open-source APMs are not strong enough to motivate app developers to choose them.

\noindent$\bullet$ \textbf{Manual instrumentation vs. Automatic instrumentation}
Two APMs (App Pulse and CA Mobile) support automatic instrumentation. Other APMs only support manual instrumentation. Based on the 500,000 apps, we find that 17 apps (<0.01\%) are instrumented with automatic instrumentation. Developers overwhelmingly choose APMs that support manual instrumentation. The reason can be twofold:

\noindent (1) Customization: Two APMs (App Pulse and CA Mobile) that support automatic instrumentation do not allows developers to make any customization, such as recording a value of a variable. In contrast, other APMs allow developers to collect runtime variable with logging functions; and

\noindent (2) Easy of use: After trying all the APMs under examination, we find that it is not hard to configure and use an APM that supports manual instrumentation. In terms of easy of use, two types of APMs is similar.

\begin{tcolorbox}[boxrule=1pt,boxsep=1pt,left=2pt,right=2pt,top=2pt,bottom=2pt]
\textbf{Popularity.} Among the 500,000 Android apps, 55,722 (11\%) apps use the APMs studied in this paper. In particular, APMs are most popular in apps from the categories, including Entertainment (52.8\%), Lifestyle (33.3\%), Game (20.1\%), and Weather (18.4\%). 

Among the 25 APMs, the top 6 popular APMs are: Flurry (32.7\%), UMeng (27.54\%), Google Firebase (15.95\%), Tencent (6.85\%), Apteligent (3.54\%), and AppFlyer (3.26\%). 

Surprisingly, we find that comparing to open source APMs, developers overwhelmingly (99.6\%) choose commercial APMs, considering the functionality, usability, costs, and even technical support.
\end{tcolorbox} 

\subsection{APMs in Practice}\label{subsec:apm-practice}

\noindent\textbf{$\bigstar$ RQ 2. Do developers still use logging frameworks (e.g., android.util.Log) even they have logging functions from APMs?}

\noindent\textbf{Motivation.} APMs provide APIs which allow developers to collect runtime data. Meanwhile, the logging functions can also be achieved by Android's built-in logging function (i.e. \code{android.util.Log}) and other logging frameworks. Therefore, we aim at understanding the intention for using both APMs' logging APIs and general logging frameworks (e.g, \code{android.util.Log}).

\noindent\textbf{Subject Logging Frameworks.} We select four most widely-used logging frameworks in Java and Android for comparison \cite{AppBrain}, including \code{android.util.Log}, \code{org.slf4j}, \code{SLF4J-android}, and \code{java.util.logging.Logger}.

\noindent\textbf{Methodology.} Similar to APM detection, we used the detection approach described in \textsection \ref{sec:methodolgy} to identify the logging frameworks used in apps. We randomly selected 100 apps that use both logging functions from APMs and logging frameworks. Note that we only consider the APMs that provide logging functions in this task. It is because we want to learn from the apps that leverage both logging frameworks and logging functions from APMs.

Similar to APM detection, we employ the detection approach described in \textsection \ref{sec:methodolgy} to identify the logging frameworks used in apps. We also record constants in logs and the type of logs (i.e., debug, warning, information). The constants in a log can present the semantic content of the log. With the type of a log, we can infer the developers' intention. Although the runtime data is no longer available for developers when an app is published, we can still extract the constants in logs and the type of logs from apps with static analysis.

\noindent\textbf{Results.} We find that 224,039 out of 500,000 apps (44.8\%) apps adopt one or more logging frameworks. 22,739 apps out of 55,722 APM-integrated apps (40.8\%) use both APM logging functions and general logging frameworks. 

\noindent\textbf{The use of logging frameworks in practice.} We collect logs and record log contents from 100 sample apps. For the contents collected in these logs, they can be categorized into four groups including \textit{string constant}, \textit{integer constants}, \textit{type of logs}, and \textit{variables}. More specifically, the \textit{string constant} and \textit{integer constants} represent the constant values used in logs. We find that developers leverage \textit{string constant} to record actions. For example, the message ``user rated the app on appirater'' is used to inform developers that users have already rated the app. The \textit{integer constants} are used to record the line numbers in source code for debugging. The \textit{variables} collected can assist developers in debugging their apps. 

By analyzing the log content collected from 100 sample apps, we find that developers leverage the logging frameworks for debugging purposes, such as recording variables for debugging, and recording source code line numbers. 

\noindent\textbf{The use of APMs in practice.} Different from using logging frameworks, developers use APM mainly for monitoring apps' performance and understanding user behaviors. To be specific, based on our observation of 100 sample apps, we learn that developers mainly use APMs for the following purposes:

\noindent$\bullet $ Monitoring the runtime performance (i.e., memory, runtime bugs, CPU usage, network performance) of their apps;

\noindent$\bullet $ Detecting network problems: Developers use APMs to diagnose potential network connection problems;

\noindent$\bullet $ User profiling: The goal of user profiling is to understand and categorize app users~\cite{Middleton:2004}. Specifically, we observe that developers mainly care about the followings: (1) understanding users' geographical distribution by inspecting their IP locations; (2) knowing users' unique hardware device IDs; (3) obtaining the device module information (e.g, Google, SAMSUNG, LG); (4) obtaining the preferences stored. Understanding users' profiles can be a double-edged sword. On the one hand, the data collected can be leveraged to improve the app, but on the other hand, it can result in privacy leakages. For example, obtaining preferences from users may violate privacy protection rules, such as GDPR \cite{GDPR,FanISSRE20}.

\noindent$\bullet $ Understanding the execution paths: Another key usage of APM is to learn how users interact with an app. A common practice is that developers instrument logs via APMs at all lifecycle methods. When a user enters or leaves a page, the APM can collect such information for developers. 

By collecting this information, developers can benefit from: (1) knowing which pages are popular among app users and which pages are seldom visited, (2) learning the time spent on each page assists developers in improving their apps, and (3) knowing the execution path can help developers infer users' behaviors and preferences.

\noindent$\bullet $ Using trace statements to monitor certain code segments: some code segments in an app require several resources (e.g., CPU and memory). Thus, monitoring the code performance at runtime is a key usage of APMs. By inspecting real-world apps, we find that developers place traces mainly for two types of code segments: time-consuming code and frequently-visited code. For the former, the execution of a code segment can be time-consuming subject to the environment, device, and other factors. For example, developers often monitor the downloading tasks (e.g., downloading a video/image from a server). The latter refers to the code segments that are executed more than once. Thus, developers place trace statements around them to monitor the performance and then optimize the code when possible.




\noindent \textbf{Use both logs and APM logging functions.} We find developers use logging frameworks even APMs provide such functions. The reason is two-fold:

\noindent$\bullet$ \textbf{Only using logging frameworks:} We find that developers cannot collect log data once their apps are released if they only using general logging frameworks. If developers intend to collect data from end users with logging frameworks, additional efforts are required. They have to set up a server and send the data collected with logs to the server. However, this can be easily achieved with APMs.

\noindent$\bullet$ \textbf{Only using APMs:} If only APMs are used in the apps, developers cannot obtain timely responses for local tasks, especially for local debugging. The data collected by APMs cannot be directly shown to developers when developers use some logging functions from APMs. It is because that most APMs do not upload the data to servers frequently.

To wrap up, the logging frameworks cannot collect runtime log data once the apps are released whereas the logging feature in APMs cannot provide timely feedback for local tasks, especially for local debugging. Consequently, there is a need for using both logging frameworks and the logging features from APMs. It also suggests that APM vendors may let their APMs support local debugging.

\begin{tcolorbox}[boxrule=1pt,boxsep=1pt,left=2pt,right=2pt,top=2pt,bottom=2pt]
\textbf{Logs vs. APMs.} 22,739 app out of 55,722 APM-integrated apps (40.8\%) also adopt logging frameworks. APMs cannot fully replace logs and vice versa. The reason is that APMs are not suitable for local debugging whereas most log frameworks cannot collect runtime data once apps are deployed. Therefore, the intentions of using APMs and logging may be different.
\end{tcolorbox} 


\noindent\textbf{$\bigstar$ RQ 3. What are the consequences of using multiple APMs in one app?}

\noindent\textbf{Motivation.} Based on our observations in RQ1, we find that app developers may use multiple APMs in one app. Specifically, 10,531 apps (out of 55,722 apps,18.9\%) use more than one APMs in apps. Therefore, we analyze how multiple APMs work at runtime. Specifically, our exploration is performed from two aspects: (\textbf{3-a}) why do developers leverage multiple APMs in one app? and (\textbf{3-b}) do multiple APMs introduce any side effect to apps?

\noindent\textbf{Methodology} To answer \textbf{3-a}, We manually inspect 100 apps that use multiple APMs in a single app. We first reverse engineer each app and then manually inspect the usage of the APMs in it. We collect and compare the following information for different APMs in a single app:

\noindent$\bullet $ The APIs leveraged by different APMs;

\noindent$\bullet $ The data collected by these APMs can be used to infer developers' intention;

To answer \textbf{3-b}, we evaluate the performance when multiple APMs are integrated into one app. Here, we select 4 APMs, including two open-source APMs (ArgusAPM \cite{ArgusAPM} and AndroidGodEye \cite{AndroidGodEye}) and two commercial APMs (Baidu and UMeng), for this task.

\noindent\textbf{Baseline experiments.}

\noindent$\bullet$ In the first baseline experiment, we evaluate whether these APMs can work properly at runtime. If one APM cannot work properly, it can affect our evaluation.

We evaluate all four APMs with the demo app respectively in the baseline experiment. We confirm that all four APMs can monitor the performance of apps at runtime. That is, they correctly offer the functionalities as they claimed, such as capturing crashes, collecting network performance, detecting ANR errors, collecting CPU and memory usages. The results can be found in Table \ref{tab:baseline-evaluation}.

\noindent$\bullet$ In the second baseline experiment, we evaluate whether a performance issue can be detected with one APM but missed when two APMs are used together.

In the first baseline experiment, we already test and confirm that all four APMs can work properly in terms of performance monitoring. Thus, we need to test whether the performance issues can still be detected under different APM pairs. Hence, we employ six APM pairs in this experiment, including <ArgusAPM, AndroidGodEye>, <ArgusAPM, BaiduAPM>, <ArgusAPM, UMengAPM>, <AndroidGodEye, BaiduAPM>, <AndroidGodEye, UMengAPM>, <BaiduAPM, UMengAPM>. After testing with different APM pairs, we find that if a performance issue can be detected with one APM, it can also be detected when two APMs are used together.

\begin{table}[!htpb]
\caption{Baseline Experiment Results.}
\centering
\begin{threeparttable}
\scalebox{0.8}{
\begin{tabular}{|c|c|c|c|c|c|}
\hline
   & Java-Native Crash & Network & ANR & CPU & Memory \\ \hline
ArgusAPM &   \cmark - \xmark  & \cmark &   \cmark    &  \cmark    &  \cmark       \\  \hline
AndroidGodEye &  \cmark - \xmark & \cmark &   \cmark    &   \cmark   &       \cmark  \\ \hline
Baidu APM &  \cmark-\cmark  & \cmark &   \cmark    &  \cmark    &  \cmark       \\  \hline
UMeng APM &  \cmark - \cmark  & \cmark &   \cmark    &   \cmark   &       \cmark  \\ \hline
\end{tabular}
}
\end{threeparttable}
\label{tab:baseline-evaluation}
\end{table}

\textbf{Multiple APMs in one app.} For two open-source APMs, we first manually instrument these two APMs by adding additional log statements to indicate the APM used. When a function in a certain APM is invoked, we are informed by the logs instrumented. Next, we integrate these APMs (i.e., open-source APM) with a self-built demo app. The demo app uses 5 components to present 5 problems, including Java-side crash, network communication error, ANR, large CPU usage, large memory consumption. However, these two open-source APMs do not cover the native crash capture function, we use two commercial APMs to capture the native crashes. Specifically, we build another demo app (demo-app-2), which embeds both Baidu APM and UMeng APM. At runtime, we trigger the native bug in the demo-app-2 to determine which APM (Baidu APM or UMeng) can capture the signal.

\noindent\textbf{Results.} 
\begin{table}[!htpb]
\caption{Multiple APM Performance}
\begin{threeparttable}
\scalebox{0.95}{
\centering
\begin{tabular}{|c|c|c|c|c|c|}
\hline
   & Crash & Network & ANR & CPU & Memory \\ \hline
First Init. APM &  \xmark(J)/\cmark(N)  & \cmark &   \cmark    &  \cmark    &  \cmark       \\  \hline
Last Init. APM &  \cmark(J)/\cmark(N)  & \cmark &   \cmark    &   \cmark   &       \cmark  \\ \hline
\end{tabular}
}
\begin{tablenotes}
    \item First Init. APM: the APM that is first initialized in the app; 
   \end{tablenotes}
\end{threeparttable}
\label{tab:mutiple-apm-study}
\end{table}

\noindent\textbf{3-a}. By checking the apps with multiple APMs, we find the following intuitions that developers use multiple APMs:

\noindent$\bullet$ \textbf{Bad programming practice.} First and foremost, it comes from the developers' poor programming practices. Developers leverage different APMs to monitor the performance of different modules in a single app. The intuition is two-fold: (1) modules are developed by different groups in a company. They use different APMs for their modules. When the entire app is built by merging modules from different groups, multiple APMs are integrated into one app; and (2) developers target at distinguishing the performance results from different parts of an app. By instrumenting different APMs for different modules, it is easy to locate the source of performance bottlenecks. However, we do not suggest such practice as it can introduce side effects (see Table \ref{tab:mutiple-apm-study}) and performance overhead (see Fig. \ref{fig:cpu-overhead}, \ref{fig:VSS-overhead}, \ref{fig:RSS-overhead}).

\noindent$\bullet$ \textbf{Combine advantages in different APMs.} Each APM has its unique features and advantages. For example, the \code{Adjust} APM provides additional functions for developers to monitor the ads in an app. The \code{UMeng} APM provides additional options for developers to monitor users' behavior under the game context. Thus, some app developers leverage different APMs with various concerns. By demystifying the APMs, we find that most functions in APMs are overlapped (see Table~\ref{tab:apm-programmingpattern}). Thus, using multiple APMs contributes less to the monitoring capability. 

\noindent\textbf{3-b}. 
To understand the side-effects introduced by the execution order of APMs, we conduct two-round testing for comparison. For the first round, we initialize ArgusAPM and then initialize AndroidGodEye in the demo app. For the second round, we test in the opposite order. The result can be found in Table \ref{tab:mutiple-apm-study}. In general, for most functions, including network diagnosis, ANR, CPU utilization, and memory usages, both of them can capture the information. However, as for the crash from the Java side, only the later initialized one can capture the crash. For native crash, we test with Baidu APM and UMeng, we find the native crash can be captured by both APMs. This is because when capturing the Java-side crash, APMs implement an \code{UncaughtExceptionHandler} to the main thread, and only the later Handler is considered as \code{UncaughtExceptionHandler} because a thread can only have one default \code{UncaughtExceptionHandler}. That is the reason why only the later initialized APM can capture the Java-side crash. However, when considering the native crash, each APM only implements a listener to capture the crash signal. That is the reason why both APMs can capture the native-side crash.

\begin{tcolorbox}[boxrule=1pt,boxsep=1pt,left=2pt,right=2pt,top=2pt,bottom=2pt]
\textbf{Consequences of using multiple APMs.}
By manually inspecting 100 apps with multiple APMs, we find that there is no sound evidence to embed multiple APMs in one app. In this study, we find that the key functions in each APM are the same even though the implementations can be different. Leveraging multiple APMs cannot improve the APMs' monitoring capability. Even worse, we find that using multiple APMs in one app can lead to side effects.
\end{tcolorbox}

\noindent\textbf{$\bigstar$ RQ 4. Will app developers collect sensitive data using APMs? }

\noindent\textbf{Motivation.} 
As APMs allow developers to collect values of variables with build-in logging functions at runtime, we aim at checking whether developers exploit APMs for stealing sensitive data from users. 

\noindent\textbf{Methodology.} We perform our experiments on the 55,722 apps with the APMs with the approach presented in Sec. \ref{subsec:usage-identification}.

\noindent\textbf{Results.} As a result, we find 23,403 apps out of 55,722 apps (42\%) collect sensitive data from users with APMs. In total, 99,583 leaks are explored in all these 23,403 apps. The top-ranked sources for these leaks are: \code{TelephonyManager::getDeviceId()} (13,943 leaks), \code{android.location.LocationManager::getLastKnownLocation()} (13,906 leaks), \code{org.apache.http.HttpEntity::getEntity} (5,030 leaks), and \code{android.location.Location::getLatitude()} (2,852 leaks). 
Next, we manually analyze data collected by developers. Then, we category them into the following groups based on their intentions. For example, \code{android.location.LocationManager::getLastKnownLocation()} and \code{android.location.Location::getLatitude()} are used to obtain the locations of app users, therefore, we categorize them into the same group based on the same intention. As a result, we find the following key intentions for developers to collect user data:

\noindent$\bullet$ Hardware device information: developers care about what kind of devices that app users adopt. Specifically, it includes (1) the unique hardware ID of a device; and (2) the module of a device (e.g., Google, SAMSUNG, LG).

\noindent$\bullet$ Users' location information: developers care about the geographic distribution of their app users;

\noindent$\bullet$ Uses' preferences: developers also care about users' preference settings in the apps. Note that, this information can be private as some private data can be embedded in the preferences; 

\noindent \textbf{Privacy leakage detection and protection.} Users can leverage our APMHunter tool to detect whether the apps collect their sensitive data via APM. Unfortunately, most app end users (e.g., elder people, children, non-IT people) do not have any CS background. It can be difficult for them to use our tool to detect malicious behaviors even if the tool is available. A more practical solution is fixing the problem from other stakeholders. We suggest two possible solutions:

Solution 1: Google Play can analyze the apps submitted by developers to identify potential privacy leaks through APMs.

Solution 2: APM vendors can limit the operations/APIs in APMs and provide non-sensitive data to developers. For example, for location data, the APMs can provide non-sensitive data (e.g., the city, the state, the country) rather than specific GPS coordinates (e.g., latitude position, longitude position).

\begin{tcolorbox}[boxrule=1pt,boxsep=1pt,left=2pt,right=2pt,top=2pt,bottom=2pt]
\textbf{Privacy.} 
It is a common practice (42\%) that apps collect private data through APMs. Developers care more about the following private information: (1) hardware device ID; (2) user location information; and (3) users' preferences. For APM vendors and app distributors (i.e. Google Play), they should carefully check the APM-based malicious behaviors.
\end{tcolorbox} 

\subsection{Performance Overhead}
\noindent\textbf{$\bigstar$ RQ 5. What is the performance overhead of using APMs?}

\noindent\textbf{Motivation.} For this RQ, we intend to discuss the additional computing resources introduced by using APMs.


\begin{figure}[!htpb]
    \centering
    \includegraphics[width=0.45\textwidth]{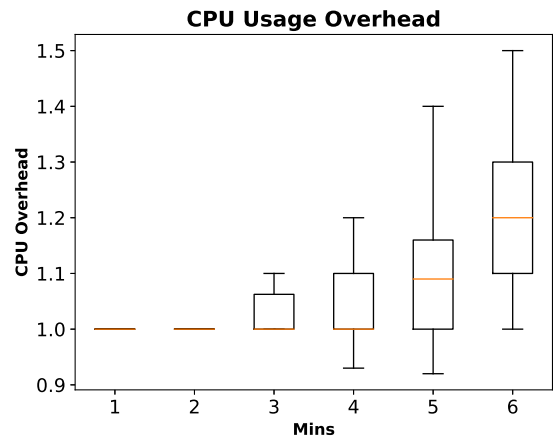}
    \caption{CPU Overhead}
    \label{fig:cpu-overhead}
\end{figure}

\begin{figure}[!htpb]
    \centering
    \includegraphics[width=0.45\textwidth]{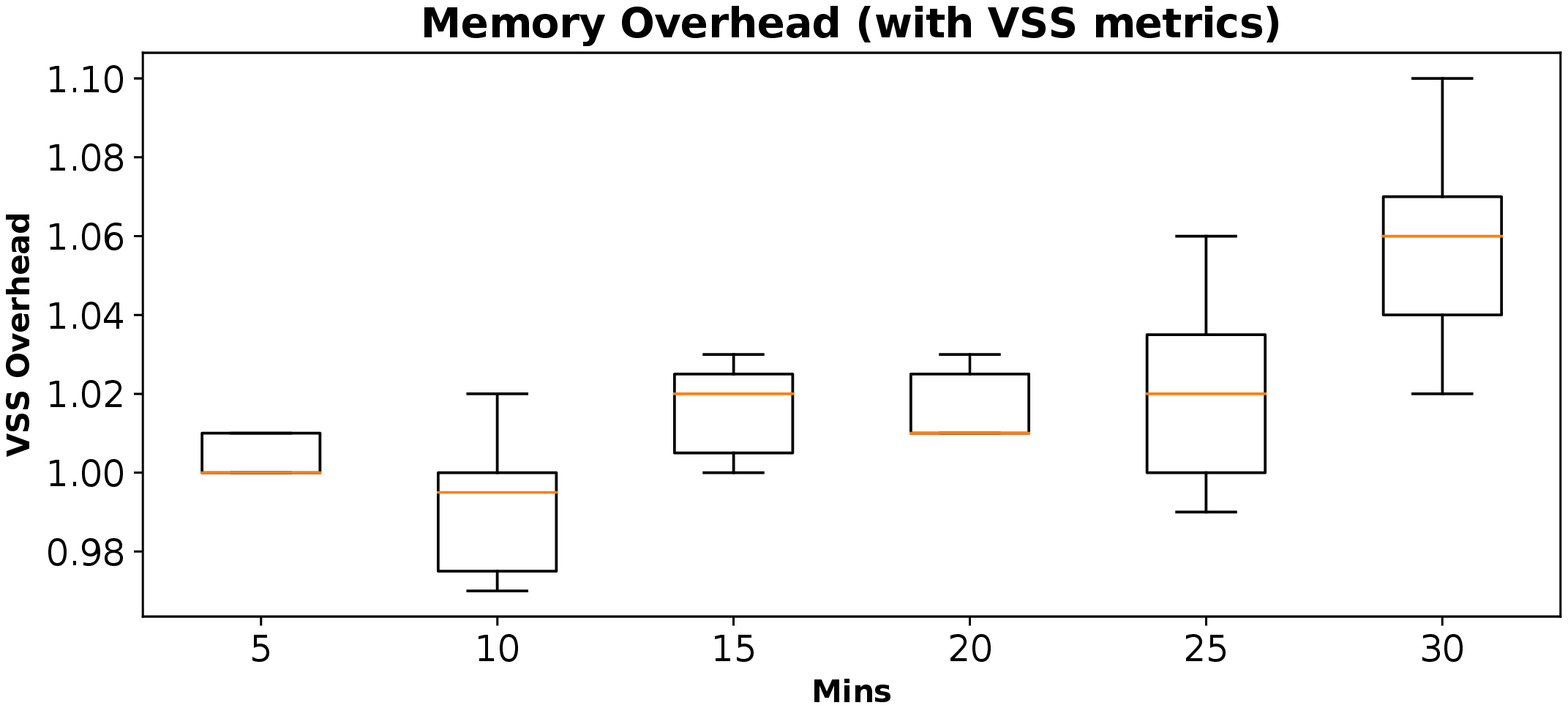}
    \caption{Memory Overhead (with VSS metrics)}
    \label{fig:VSS-overhead}
\end{figure}

\begin{figure}[!htpb]
    \centering
    \includegraphics[width=0.45\textwidth]{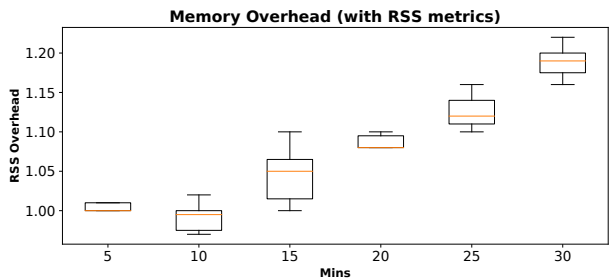}
    \caption{Memory Overhead (with RSS metrics)}
    \label{fig:RSS-overhead}
\end{figure}

\noindent\textbf{Environment.} We randomly select 50 apps that each of them uses only one APM under study. These apps are installed on our test machine, which is a Pixel phone. The Android version of the Pixel phone is Android P (API 27).

\noindent\textbf{Methodology.} 
We develop a tool to repackage an APK by removing all APM API invocations from the app. As a result, for each APK, we obtain a repackaged version that does not contain any APM usage. To evaluate the performance overhead introduced by APMs, we compare these two versions (with APMs and without APMs) using  Sapienz~\cite{Mao:2016}, which is a widely used fuzzing analysis tool for Android apps. We test each app using Sapienz for half an hour. During the testing, the memory consumption of an app is calculated with \code{top} utility tool in adb (\code{adb shell top}) and we further resort to \code{adb shell dumpsys cpuinfo} to get the CPU usage information. We set the sampling interval to 10 seconds.

\noindent\textbf{Results.}
%
Fig. \ref{fig:cpu-overhead} shows CPU usage overhead with the box plot. The x-axis represents the total minutes that the app has been executed. The y-axis represents the overheads. For example, node (5,1.01) represents that after 5 minutes, the CPU usage of the app with APM is 1.01 times higher than the app without APM. The average values of CPU overhead can be 1 to 1.2 times higher than that of apps without APMs. The median values of CPU overhead range from 1 to 1.2. As for memory consumption, two common metrics are adopted in Unix-like systems (e.g., Android). The Resident Set Size (RSS) shows how much memory is currently used by the process  (i.e., the app in our context). While the Virtual Set Size (VSS) shows how much memory is allocated to the process. Fig. \ref{fig:VSS-overhead} and \ref{fig:RSS-overhead} show the memory overhead with VSS and RSS metrics respectively. Specifically, as for VSS usage, the costs for apps with APMs can be 1 to 1.058 times higher than that of apps without APMs. The median values of VSS range from 1 to 1.06. As for RSS usage, the costs for apps with APMs can be 1 to 1.189 times higher than that of apps without APMs. The median values of RSS range from 1 to 1.192.

In summary, even though using an APM in an app requires additional computational resources, the overhead introduced is not high.

\begin{tcolorbox}[boxrule=1pt,boxsep=1pt,left=2pt,right=2pt,top=2pt,bottom=2pt]
\textbf{Performance Overhead.} APMs introduces overheads in terms of memory and CPU usage. By conducting an experiment on 50 apps, we find that the overhead introduced by APMs is not high.
\end{tcolorbox}

%% file: sections/lessonslearned.tex
\section{Lessons Learned}\label{sec:lessons-learned}
In this section, we summarize our findings in this study and provide suggestions and tips for stakeholders.

\subsection{Suggestions for APM Vendors}
\noindent\textbf{(1) Avoid requesting dangerous or deprecated permissions.}
As summarized in Sec. \ref{subsec:limitations-drawbacks}, several APMs request deprecated or even dangerous permissions, such as READ\_LOGS, READ\_PHONE\_STATE, from app users. These permissions are officially marked as dangerous and deprecated by Android \cite{Android}.

Using deprecated permissions can cause some functions in APMs to be no longer valid/supported in the latest Android versions, which can lead to potential compatibility issues. Even worse, using deprecated permissions in apps can result in critical security issues. For example, the work \cite{fratantonio17} shows that the attacks can be launched once the permission \code{SYSTEM\_ALERT\_WINDOW} is authorized. 

\noindent\textbf{(2) Avoid accessing sensitive files.}
As presented in Sec.\ref{subsec:limitations-drawbacks}, some files (e.g.,\code{proc/stat}) in the Android system store the sensitive data. APMs must be prohibited from accessing these files. Inappropriate usage of APMs can cause privacy leaks. As introduced in Sec. \ref{sec:apm}, some optional solutions can be adopted by vendors to prevent the use of these sensitive files. For example, they can access the file \code{/proc/cpuinfo} rather than the file \code{proc/stat} for CPU usage.

\noindent\textbf{(3) Introduce additional features for handing known performance anti-patterns.} As presented in Sec.\ref{subsec:confront-anti-patterns}, we summarize 8 common performance anti-patterns from existing studies \cite{Liu:2014,Afjehei:2019,Chen:2018,Hecht:2015} to evaluate whether existing APMs are suitable for resolving them. Unfortunately, existing APMs cannot diagnose all anti-patterns. Sometimes, additional human efforts are required to explore performance bottlenecks. Therefore, we suggest APM vendors extend the functionalities of APMs to support all common performance anti-patterns aforementioned.

\noindent\textbf{(4) Respecting to the changes on the Android System.} We find that APM vendors update their APMs without considering the changes on Android (see. Sec. \ref{subsec:apm-evolution}). However, the changes in the Android system can influence the performances of APMs. For example, APMs can access file \code{/proc/stat} for collecting CPU information. However, since Android 8.0 (API 26), such access is prohibited. Therefore, we suggest that APM vendors respect the changes in the Android system when updating their products.

\noindent\textbf{(5) Privacy management.} When an app is integrated with an APM, the users' runtime data is collected by the APM. Users cannot terminate such collection. From users' perspectives, they should be able to determine whether they would like to share their data. Thus, APM vendors should provide such options.

\noindent\textbf{(6) Preventing app developers from building malicious apps with APMs.} Even though it is not a common practice to leverage APMs in malicious apps, we still find that some malicious apps use APMs to collect users' passwords, addresses, and so forth. Thus APM vendors must carefully monitor the data collected through APMs. 

\subsection{Suggestions for App Developers}
\noindent\textbf{(1) Avoid collecting private data from users through APMs.}
As presented in Sec.\ref{subsec:apm-practice}, 42\% of apps leverage APMs to collect private data from users. The data collected includes device ID, location information, device modules, and so forth. Malicious apps can collect users' private data through their custom code. The existing tools/studies \cite{Zhou:2012,Li:2017,Malton17,xue2019ndroid} detect malicious apps by inspecting their code or execution traces. However, our study finds that malicious apps can collect private data through APMs. 

Even though data from users can assist developers in improving their apps, developers must carefully inspect the data collected through APMs. For example, developers can collect coarse-granularity location data instead of fine-granularity location data. One guidance that developers can refer to is the General Data Protection Regulation of EU (EU GDPR) \cite{GDPR}. Moreover, developers should clarify the data collected in their apps' privacy policies \cite{PPCheckerTSE19,AutoPPG,YLTSE18,Yu:DSN:2016}.

\noindent\textbf{(2) Using one APM rather than more.}
As presented in Sec.\ref{subsec:apm-practice}, using more than one APM, for most times, cannot contribute to app monitoring and information collection. However, more APMs may introduce additional side effects to apps, such as additional costs in CPU and memory. Besides, some properties of apps are collected more than once (e.g., CPU, memory, and ANRs). But the crash can only be capture by the latest initialized APM. Even though each APM has its unique features and advantages, we do not suggest developers for using more than one APM in an app. For example, some APMs can help developers detect whether the ads in apps are displayed at runtime. Some APMs allow developers to record how users are interacting with their game with ease. Based on our investigation, it is not hard to manually implement these features. For example, to detect whether the ads in apps are displayed at runtime, developers can leverage the logging functions in the APM. Therefore, developers can implement the feature needed on their own rather than importing another APM. We recommend developers to remove redundant APMs from their apps.

%% file: sections/relatedwork.tex
\section{Related Work}\label{sec:relatedwork}
\subsection{Application Performance Monitoring} 
Trubiani et al.'s work \cite{Trubiani:2018} discussed how to use the data collected by APM to diagnose the network bottleneck in applications. Ahmed et al.'s work \cite{Ahmed:2016} studied the effectiveness of APMs for measuring the runtime performance of web applications. Yao et al.'s work \cite{Yao:2018} discussed the way to improve the performance of system monitoring by instrumenting logs. Willnecker et al. \cite{Willnecker:2015} proposed an approach to model the performance of JavaEE applications with APMs. Different from these studies, we focus on exploring the functionalities of Android-oriented APMs and discovering the usage of APMs in real-world apps instead of the ways to use the data collected by APMs.

\subsection{Measurement and Monitoring for Apps}

\noindent\textbf{Network Measurement.} Since the Android framework provides convenient interfaces for users to intercept and forward network packets, many Android apps~\cite{MopEye:2017} target for measuring mobile network performance. To scrutinize the measurement accuracy of these apps, many studies have been proposed. Li et al.~\cite{Li:2015} adopted the network round-trip time (nRTT) as the metric to appraise the accuracy of network measurement apps, and they found that nRTTs measured by these apps are heavily inflated. Xue et al.~\cite{Xue:2017} conducted a systematic study on three types of factors, including implementation patterns of measurement apps, Android architecture, and network protocols, to learn how these factors influence the measurement results of these apps. Li et al.~\cite{Li:2018} pointed out that the delay inflation, which influences the accuracy of network measurement apps, can be introduced from (1) the long path of sub-function invocations in Android runtime; and (2) the sleeping functions in the drivers between the kernel and physical layer.

\noindent\textbf{App Monitoring.} To diagnose performance bottlenecks in apps, several approaches have been proposed to conduct efficient app monitoring. AppInsight~\cite{AppInsight:2012} instrumented mobile apps by interposing event handlers to collect information on critical paths that are triggered by user transactions. Lee et al.~\cite{Lee:2014} proposed a user interaction-based mobile application profiling system, which analyzes fine-grained information, including user interaction, system behavior, and power consumption, to perform Android app tuning. AndroidPerf~\cite{AndroidPerf:2015}, a cross-layer profiling system, leverages cross-layer dynamic taint analysis and instrumentation to obtain both the execution information and the performance information about Android apps. DiagDroid~\cite{DiagDroid:2016} adopted a dynamic instrumentation approach, which is based on abstractions of various categories of UI-triggered asynchronous executions, to capture the data related to UI interactions and diagnose UI performance of apps. Technically, all these works concentrate on app monitoring by instrumenting the subject apps rather than leveraging APMs.

\subsection{Instrumentation}
Instrumentation is another frequently used technique in program understanding, debugging and testing\cite{Karami:2013,Backes:2017,Arzt:2014,Mao:2016,Gu:2017}. Similar to APMs, developers sometimes use logs to collect information and debug apps. Karami et al.~\cite{Karami:2013} proposed an approach that uses instrumentation to analyze and model the behavior of an app. Specifically, it focused on file I/O and network connection issues by inspecting API calls. Similarly, the tool ARTist assists developers in understanding program behaviors by extracting APIs and arguments in the program \cite{Backes:2017}. The goal of ARTist is to help developers understand the program and reveal the malicious behaviors in the program. Other studies adopted instrumentation techniques to test Android apps for diagnosing bugs and monitoring performance \cite{Arzt:2014,Mao:2016,Gu:2017}. 

%% file: sections/conclusion.tex
\section{Conclusion}
\label{sec:conclusion}
Although more and more apps adopt APMs, developers use them as black-box tools. In this paper, we demystify the functionalities of APMs in detail. We discuss 9 commons functions in APMs, and we reveal 7 design defects in APMs. We also evaluate the performance side-effects introduced by using APMs. To evaluate and study how APMs are used in practice, we developed a prototype named APMHunter. We evaluate the performance of APMHunter on 500 apps from Google Play. The results show that (1) in terms of detecting APMs in apps, the precision of APMHunter is 97.7\% and the recall is 87.5\%; (2) in terms of identifying sensitive UI elements in apps, the precision of APMHunter is 95.5\% and the recall is 94.1\%; and (3) as for identifying privacy leaks, the precision of APMHunter is 96.2\% and the recall is 84.6\%. Our large-scale empirical study on 500,000 Android apps implies that 23,403 apps collect sensitive data with APMs. Therefore, we suggest both app developers and APM vendors need to vigilant in protecting users' private data.


%% file: bare_jrnl.bbl
\begin{thebibliography}{10}
\bibitem{appdevsurvey1}
Techbencon, ``Performance engineering survey: Findings from 400 dev, test, and
  it ops professionals.''
\end{thebibliography}
